\documentclass{aa}
\usepackage{graphicx}
\usepackage[figuresright]{rotating}		% sidewaysfigure, etc.
  \usepackage[authoryear]{natbib}	        % Bibliographie-Paket
    \usepackage{rotating}
  \usepackage{subfigure} 
  \usepackage{tabularx}
\usepackage{txfonts}

\usepackage{longtable,lscape}
\begin{document}
   \title{Effects of primordial chemistry on the cosmic microwave background}

%   \subtitle{I. Overviewing the $\kappa$-mechanism}

   \author{Dominik R. G. Schleicher
          \inst{1}
\and
         Daniele Galli
         \inst{2}
\and
         Francesco Palla
          \inst{2}
          \and
          Max Camenzind\inst{3}
\and
        Ralf S. Klessen
           \inst{1}
\and 
      Matthias Bartelmann
                  \inst{1}
\and
       Simon C. O. Glover
            \inst{4}
          }

   %\offprints{D. Schleicher}

   \institute{Institute of Theoretical Astrophysics / ZAH,
              Albert-Ueberle-Str. 2, D-69120 Heidelberg, Germany\\
              \email{dschleic@ita.uni-heidelberg.de, rklessen@ita.uni-heidelberg.de, mbartelmann@ita.uni-heidelberg.de}
\and
INAF-Osservatorio Astrofisico di Arcetri, Largo E. Fermi 5, I-50125 Firenze, Italy\\
\email{galli@arcetri.astro.it, palla@arcetri.astro.it}
         \and
             Landessternwarte Heidelberg / ZAH, Koenigstuhl 12, D-69117 Heidelberg, Germany\\
             \email{mcamenzind@lsw.uni-heidelberg.de}
\and
           Astrophysikalisches Institut Potsdam, An der Sternwarte 16, D-14482 Potsdam, Germany\\
               \email{sglover@aip.de}
             }

%\date{Received September 15, 1996; accepted March 16, 1997}

% \abstract{}{}{}{}{} 
% 5 {} token are mandatory

\newcommand{\HI}{$\mathrm{H}$\ }
\newcommand{\HII}{$\mathrm{H}^+$\ }
\newcommand{\HzII}{$\mathrm{H}_2^+$\ }
\newcommand{\HzI}{$\mathrm{H}_2$\ }
\newcommand{\HM}{$\mathrm{H}^-$\ }
\newcommand{\HeM}{$\mathrm{He}^-$\ }
\newcommand{\HeHII}{$\mathrm{HeH}^+$\ }
\newcommand{\HeI}{$\mathrm{He}$\ }
\newcommand{\HeII}{$\mathrm{He}^+$\ }
\newcommand{\HeIII}{$\mathrm{He}^{++}$\ }
\newcommand{\DI}{$\mathrm{D}$\ }
\newcommand{\DII}{$\mathrm{D}^+$\ }
\newcommand{\HDII}{$\mathrm{HD}^+$\ }
\newcommand{\HDI}{$\mathrm{HD}$\ }
\newcommand{\DM}{$\mathrm{D}^-$\ }
\newcommand{\e}{$\mathrm{e}$\ }

\newcommand{\HId}{$\mathrm{H}$}
\newcommand{\HIId}{$\mathrm{H}^+$}
\newcommand{\HzIId}{$\mathrm{H}_2^+$}
\newcommand{\HzId}{$\mathrm{H}_2$}
\newcommand{\HMd}{$\mathrm{H}^-$}
\newcommand{\HeMd}{$\mathrm{He}^-$}
\newcommand{\HeHIId}{$\mathrm{HeH}^+$}
\newcommand{\HeId}{$\mathrm{He}$}
\newcommand{\HeIId}{$\mathrm{He}^+$}
\newcommand{\HeIIId}{$\mathrm{He}^{++}$}
\newcommand{\DId}{$\mathrm{D}$}
\newcommand{\DIId}{$\mathrm{D}^+$}
\newcommand{\HDIId}{$\mathrm{HD}^+$}
\newcommand{\HDId}{$\mathrm{HD}$}
\newcommand{\DMd}{$\mathrm{D}^-$}
\newcommand{\ed}{$\mathrm{e}$}

\newcommand{\fHI}{\mathrm{H}}
\newcommand{\fHII}{\mathrm{H}^+}
\newcommand{\fHzII}{\mathrm{H}_2^+}
\newcommand{\fHzI}{\mathrm{H}_2}
\newcommand{\fHM}{\mathrm{H}^-}
\newcommand{\fHeM}{\mathrm{He}^-}
\newcommand{\fHeHII}{\mathrm{HeH}^+}
\newcommand{\fHeI}{\mathrm{He}}
\newcommand{\fHeII}{\mathrm{He}^+}
\newcommand{\fHeIII}{\mathrm{He}^{++}}
\newcommand{\fDI}{\mathrm{D}}
\newcommand{\fDII}{\mathrm{D}^+}
\newcommand{\fHDII}{\mathrm{HD}^+}
\newcommand{\fHDI}{\mathrm{HD}}
\newcommand{\fDM}{\mathrm{D}^-}
\newcommand{\fe}{\mathrm{e}}
\newcommand{\fpp}{\mathrm{p}}

\abstract {Previous works have demonstrated that the generation of secondary CMB anisotropies due to the molecular optical depth is likely too small to be observed. In this paper, we examine additional ways in which primordial chemistry and the dark ages might influence the CMB.}
{We seek a detailed understanding of the formation of molecules in the postrecombination universe and their interactions with the CMB. We present a detailed and updated chemical network and an overview of the interactions of molecules with the CMB.}
{We calculate the evolution of primordial chemistry in a homogeneous universe and determine the optical depth due to line absorption, photoionization and photodissociation, and estimate the resulting changes in the CMB temperature and its power spectrum. Corrections for stimulated and spontaneous emission are taken into account.}
{The most promising results are obtained for the negative hydrogen ion \HM and the \HeHII molecule. The free-free process of \HM yields a relative change in the CMB temperature of up to $2\times10^{-11}$, and leads to a frequency-dependent change in the power spectrum of the order $10^{-7}$ at 30 GHz. With a change of the order $10^{-10}$ in the power spectrum, our result for the bound-free process of \HM is significantly below a previous suggestion. \HeHII efficiently scatters CMB photons and smears out primordial fluctuations, leading to a change in the power spectrum of the order $10^{-8}$.}
{We demonstrate that primordial chemistry does not alter the CMB during the dark ages of the universe at the significance level of current CMB experiments. We determine and quantify the essential effects that may contribute to changes in the CMB and leave an imprint from the dark ages, thus constituting a potential probe of the early universe.}

   \keywords{molecular processes --
                atomic processes --
                cosmology: early universe -- cosmic microwave background -- observations -- theory .
               }

   \maketitle
%
%________________________________________________________________

\section{Introduction}
The cosmic microwave background (CMB) is one of the most powerful tools of high-precision cosmology, as it allows one to determine the cosmological parameters, the power spectrum of initial fluctuations and various other quantities. It is thus important to have a detailed theoretical understanding of all effects that have a potential influence on CMB measurements. Following the WMAP 3 year and 5 year results \citep{Spergel,Hinshaw, Komatsu08} that confirmed our standard picture of cosmology, we are looking forward to the precise measurement that will be performed with Planck\footnote{http://www.rssd.esa.int/index.php?project=planck} in only a few years. The measurement of the electron scattering optical depth allows one to constrain the effective reionization redshift and yields indirect information about an epoch that cannot yet be observed. Recalling that the cross sections of bound electrons can be larger by orders of magnitude compared to the cross section of free electrons, the optical depth due to molecules may provide information on the dark ages of the universe, in spite of the small molecular abundances. In fact, the early work of \citet{Maoli94} suggested that the molecular opacities could smear out CMB fluctuations on the scale of the horizon, and at the same time create new secondary fluctuations due to the interaction with the velocity fields which are present in proto-clouds in the dark ages. This work had the intention to explain why no CMB fluctuations had been observed at that time. \\ \\
Since then, there has been considerable progress both in the understanding of chemical processes in the early universe and the molecular abundances, as well as in the interaction of molecules with the CMB and the generation of spectral-spatial fluctuations. While recombination was originally examined by \citet{PeeblesRec} and \citet{ZeldovichRec}, and improved in several follow-up works \citep{Matsuda, Jones, Sasaki} based on analytic methods, today's computers allow a detailed treatment of the recombination process based on a reaction network that takes into account hundreds of energy levels for \HId, \HeI and \HeIId, as in the work of Seager, Sasselov and Scott \citep{Seager}. A simplified code reproducing the results of this detailed calculation is given by Seager, Sasselov and Scott \citep{SeagerFast}.  In a recent series of papers, \citet{Switzer1, Switzer2, Switzer3} considered the recombination of helium in great detail. Deviations of the CMB spectrum from a pure blackbody have also been considered in various works. \citet{Dubrovich75} considered the effect of hydrogen recombination lines. \citet{Rubino2, Rubino1} examined distortions due to helium and hydrogen lines in more detail, and \citet{Chluba} examined distortions due to the two-photon process. \citet{Sunyaev} describe the rich physics involved in the recombination process in great detail, and \citet{Wong} give a good overview of recent improvements and uncertainties regarding the recombination process. \\ \\

Here, the main focus is on the postrecombination universe and possible imprints in the CMB from this period. The formation of \HzI during the dark ages has already been discussed by \citet{Saslaw}. A more detailed treatment of molecules has been performed by \citet{Puy}, Stancil, Lepp and Dalgarno \citep{Stancil} and by Galli and Palla \citep{Galli}. A useful collection of analytic formulae for estimating the abundance of various molecules after recombination was given by Anninos and Norman \citep{AnninosIn}. Recently, this problem was re-examined by Puy and Signore \citep{Puynew}, and Hirata and Padmanabhan \citep{Hirata} examined \HzI formation in more detail, taking into account the effects due to non-thermal photons. \\ \\
Regarding the interaction of molecules with the CMB, the effects of various molecules due to their optical depths have been considered by \citet{Dubrovich1994}. The enhancement of spectral-spatial fluctuations due to the luminescence effect, which is well-known from stars in reflection nebulae, has also been discussed by \citet{Dubrovich1997}. Observational prospects for Herschel and ODIN have been discussed by \citet{Maoli05}, and the relevance for Planck has been assessed by \citet{Dubrovich2007}. The effects discussed were based on smear-out of primary CMB fluctuations due to the molecular optical depth and the generation of secondary anisotropies due to scattering with proto-objects in the universe. \citet{Mayer} provided a recent overview of different processes that may contribute to the opacity in primordial gas, and derived Rosseland and Planck mean opacities. \\ \\
\citet{Black} recently considered the influence of the bound-free transition of the negative hydrogen ion on the CMB, and found an optical depth of more than $10^{-5}$ at $10\ \mathrm{cm}^{-1}$, which would have interesting implications for Planck and other CMB experiments. This contributed to our original motivation to examine this and other effects in more detail. As we will show below, however, the optical depth is due to $\fHM$ bound-free transitions is much smaller than reported by \citet{Black}, and must in addition be corrected for stimulated and spontaneous emission. On the other hand, there are other effects from \HM and \HeHII that are close to observational relevance.\\ \\

From a numerical point of view, the pioneering work of \citet{Anninos} provided a flexible and easily extendible scheme that is still widely used in state-of-the art simulations of the early universe, and which is also adopted in the public version of the Enzo code \citep{O'SheaEnzo, Bryan, Norman}\footnote{http://lca.ucsd.edu/portal/software/enzo}. This scheme is extended here to account for effects during recombination and the evolution of primordial chemistry in the homogeneous universe. We further provide an extended overview of the essential processes that may influence the CMB, determine the contributions from the most relevant species, and discuss the possibilities for detection with the Planck satellite. In section \ref{Recombination}, we present the general picture regarding the formation of the first molecules in the universe and give some analytic estimates for the abundances. In section \ref{effects}, we provide a general discussion of the potential effects of primordial chemistry on the CMB. In section \ref{Network}, we present the chemical network for our calculation, which includes some new rates for the \HeHII molecules that are given in the appendix. Section \ref{Enzo} explains the numerical algorithm for the chemical network, and section \ref{Results} presents the model abundances as a function of redshift. In section \ref{CMB}, we explain our treatment of the different species and discuss the observational implications. Further discussion and outlook is provided in section \ref{discussion}.

\section{Imprints from primordial molecules on the CMB}\label{effects}
The interaction with the CMB is crucial to determine the evolution and abundance of primordial molecules, and conversely, this interaction may also leave various imprints on the CMB photons while they travel through the dark ages. The most obvious imprint on the CMB is probably a frequency-dependent change in the observed CMB spectrum $I(\nu)$ due to absorption by a species $M$. An upper limit on this effect is given by
\begin{eqnarray}
I(\nu)=B(\nu)e^{-\tau_M(\nu)},\label{cmbobs}
\end{eqnarray}
where $B(\nu)$ denotes the unaffected CMB spectrum and the optical depth $\tau_M(\nu)$ of species $M$ at an observed frequency $\nu$ is given by an integration over redshift as
\begin{eqnarray}
\tau_M(\nu)&=&\int dl\ \sigma_M\left[\nu_0(1+z)\right] n_M  \label{taumol} \\
&=& \frac{n_{H,0}c}{H_0}\int_{0}^{z_f} f_M(z)\sigma_M\left[\nu_0(1+z)\right]\frac{(1+z)^2}{\sqrt{\Omega_\Lambda+(1+z)^3\Omega_m}}dz,\nonumber
\end{eqnarray}
where $\sigma_M(\nu)$ is the absorption cross section of the considered species as a function of frequency, $n_M$ the number density of the species, $n_{H,0}$ the comoving hydrogen number density, $c$ the speed of light, $H_0$ Hubble's constant, $f_M$ the fractional abundance of the species relative to hydrogen, and $z_f$ the redshift at which it starts to form efficiently. Such an optical depth can be provided by resonant line transitions from molecules with high dipole moments like \HeHII and \HDIId, free-free processes or photodestruction of species like \HMd, \HeM or \HeHIId, which have a relatively low photodissociation threshold. However, Eq.~\ref{cmbobs} gives only an upper limit because absorption can be balanced by inverse processes (spontaneous and
stimulated emission). A better estimate can be obtained by introducing an excitation temperature $T_{\rm ex}$ defined by
\begin{equation}
\frac{n_u}{n_l}=\frac{g_u}{g_l}\mathrm{exp}\left(-\frac{E_u-E_l}{kT_{ex}} \right)
\end{equation}
where $n_u$ and $n_l$ denote the population of the upper and the lower level, $g_u$ and $g_l$ are the corresponding statistical weights and $k$ is Boltzmann's constant. The excitation temperature is essentially determined by the ratio of collisional and radiative de-excitations. For molecules with non-vanishing dipole moments, the de-excitation is dominated by radiative transitions, and the level populations are in equilibrium with the radiation temperature. The frequency-dependent change in the radiation temperature is given by
\begin{equation}
\Delta T_r\sim -\left(T_r-T_{\rm ex}\right)\tau_M(\nu).
\end{equation}
This expression indicates a major obstacle for the detection of primordial molecules: those species with high dipole moments and thus high cross sections have an excitation temperature which is very close to the temperature of the CMB, while species with low dipole moments have a very low optical depth. Molecular resonant line transitions are thus unlikely to lead to a non-negligible net change in the radiation temperature. For photodestruction processes, this is different because the destruction is regulated by the CMB temperature, while the inverse formation processes are governed by the gas temperature. Thus, photodestruction can lead to a net change in the number of CMB photons. The same is true for free-free processes, which emit a blackbody spectrum according to the gas temperature, but absorb the spectrum of the radiation field. \\ \\
In addition, as discussed by \citet{Maoli94}, the optical depth from primordial molecules may smear out primary fluctuations in the CMB if the optical depth acts in a way that effectively scatters the CMB photons. This is possible even in a situation where stimulated and spontaneous emission balance the absorption of molecules, such that there is no net change in the number of photons. It is thus somewhat complementary to the effect discussed above. However, it must be noted that it requires spontaneous emission to be important, as stimulated emission does not change the direction of the photons, and does not provide a mechanism for scattering. We thus emphasize the importance to correct for stimulated emission. As shown by \citet{Basu04} for small angular scales, such an optical depth (corrected for stimulated emission) then leads to a change in the power spectrum given by 
\begin{equation}
\Delta C_l\sim-2\tau C_l,\label{changepower}
\end{equation}
where the $C_l$'s are the usual expansion coefficients for the observed power spectrum. As discussed by \citet{Maoli94}, such scattering processes can in addition generate secondary anisotropies which are proportional to the optical depth of the scattering processes. However, this effect is suppressed by more than three orders of magnitude, as it is also proportional to the ratio of the peculiar velocity to the speed of light.  \citet{Dubrovich1997} discussed the luminescence effect of various molecules which could potentially amplify the generation of secondary anisotropies. Based on the new abundances found in this work, we will give a basic estimate of this effect in section \ref{discussion}. \citet{Basu07} considered in addition the effects of emission from molecules like \HDI and $\mathrm{LiH}^+$, as \HDI in particular is an important coolant in cold primordial gas. Unfortunately, the effect seems to be negligible.

\section{Recombination and the formation of molecules in the early universe}\label{Recombination}
\subsection{Hydrogen recombination}
\begin{figure}\begin{center}
{\hspace{-0.25cm}}
\includegraphics[scale=0.45]{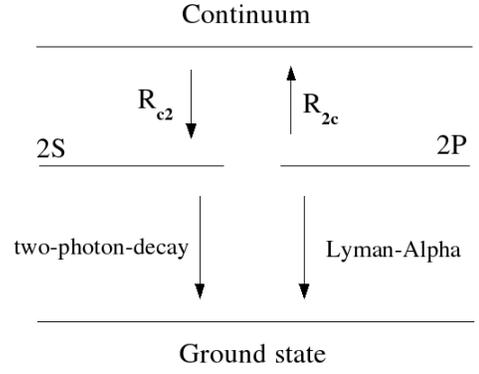}
\caption{The idealized three-level hydrogen atom (ground state $1s$, excited states $2s$ and $2p$, continuum), and the relevant transitions.}
\label{img:recomb}\end{center}
\end{figure}
Various approaches exist to calculate the time evolution of the ionized hydrogen fraction during recombination. The traditional approach is based on solving one single ordinary differential equation for the ionization degree $x_\fe$, which can be done using the approximation of an effective three-level-atom. This was first done by \citet{PeeblesRec}, and an improved version of this equation was derived by \citet{Jones}. This treatment was based essentially on the following assumptions:
%\begin{eqnarray}
%-\frac{dx_e}{dt}&=&[x_e^2n_H R_{c2}\left\{1+K\Lambda n_H(1-x_e)+KR_{c2}n_H^2x_e^2 \right\}\nonumber\\
%&-&R_{2c}(1-x_e)e^{-h\nu_\alpha/k_BT_r}
%\{ 1+K\Lambda n_H(1-x_e)\}]\nonumber\\&\times&[1+K\Lambda n_H(1-x_e)+KR_{2c}n_H(1-x_e)\nonumber\\
%&+&KR_{c2}x_e^2n_H^2]^{-1}.
%\end{eqnarray}
%Here, $\nu_\alpha$ is the Lyman $\alpha$ frequency, $T_r$ the temperature of radiation, $\Lambda=8.227\mathrm{ s}^{-1}$ the two-photon-decay from the first excited state, $R_{c2}$ the rate of radiative transitions from the continuum to the excited states, $R_{2c}$ the rate of radiative transitions from the excited states to the continuum, and $K=(c^3/8\pi\nu_\alpha^3)(a/\dot{a})$ is proportional to the expansion timescale of the universe with scale-factor a. The basic physical processes are illustrated in Fig. \ref{img:recomb}. The equation is based on the following assumptions:
\begin{itemize} 
 \item It is sufficient to take into account only hydrogen (helium is not considered).
 \item Collisional processes are outweighted by radiative processes and may be ignored.
 \item The relative populations of the excited fine structure states are thermal.
 \item Recombination and photoionization rates to and from higher states are related by Saha's formula.
 \item Any such recombination cascades down to settle in the first excited state.
 \item The level populations obey $n_{1s}\ll n_{2s}$, with $n_{1s}$ the number density of hydrogen atoms in the ground state and $n_{2s}$ the corresponding density for the 2s state.
 \item Each net recombination gives rise to a Lyman-$\alpha$ photon or two others of lower energy.
  \end{itemize}
  
\citet{Seager} presented a very detailed calculation independent of these a-priori assumptions, treating helium and hydrogen as multi-level atoms with several hundred levels and evolving one ordinary differential equation (ODE) for each level, using a self-consistent treatment of the radiation field and its interaction with matter, which effectively leads to a speed-up of recombination. To allow the integration of their new method in other cosmological applications like CMBFAST \citep{CMBFAST}, \citet{SeagerFast} produced a simplified version of this code which is capable of reproducing the results of the full multilevel treatment. This simplified code, RECFAST\footnote{http://www.astro.ubc.ca/people/scott/recfast.html}, solves three ODEs for the ionized hydrogen fraction $x_\fpp$, the ionized helium fraction $x_{\fHeI}$ and the gas temperature T. These ODEs are parametrized so as to reproduce the speed-up in the recombination process. In this paper, we make use of the ODE solving for the ionized hydrogen fraction, which is given by
\begin{eqnarray}
\frac{dx_\fpp}{dz}&=&\frac{[x_\fe x_\fpp n_\fHI \alpha_\fHI -\beta_\fHI (1-x_\fpp)e^{-h_p\nu_{\fHI, 2s}/kT}] }{H(z)(1+z)[1+K_\fHI(\Lambda_\fHI+\beta_\fHI)n_\fHI(1-x_\fpp)]}\nonumber\\
&\times&[1+K_\fHI\Lambda_\fHI n_\fHI(1-x_\fpp)].
\end{eqnarray}
In this equation, $n_\fHI$ is the number density of hydrogen atoms and ions, $h_p$ Planck's constant and the parametrized case B recombination coefficient for atomic hydrogen $\alpha_H$ is given by
\begin{equation}
\alpha_\fHI=F\times10^{-13}\frac{at^b}{1+ct^d}\ \mathrm{cm}^3\ \mathrm{s}^{-1}
\end{equation} 
with $a=4.309$, $b=-0.6166$, $c=0.6703$, $d=0.5300$ and $t=T/10^4\ K$, which is a fit by \citet{Pequignot} to the coefficient of \citet{Hummer}. This coefficient takes into account that direct recombination into the ground state does not lead to a net increase of neutral hydrogen atoms, since the photon emitted in the recombination process can ionize other hydrogen atoms in the neighbourhood.

 The fudge factor $F=1.14$ serves to speed up recombination and is determined from comparison with the multilevel-code. The photoionization coefficient $\beta_\fHI$ is calculated from the recombination coefficient as $\beta_\fHI=\alpha_\fHI(2\pi m_\fe kT/h_p^2)^{3/2}\mathrm{exp}(-h_p\nu_{\fHI, 2s}/kT)$. The wavelength $\lambda_{\fHI, 2p}$ corresponds to the Lyman-$\alpha$ transition from the $2p$ state to the $1s$ state of the hydrogen atom. The frequency for the two-photon transition between the states $2s$ and $1s$ is close to Lyman-$\alpha$ and is thus approximated by $\nu_{\fHI, 2s}=c/\lambda_{\fHI, 2p}$, where $c$ is the speed of light (i.e., the same averaged wavelength is used). Finally, $\Lambda_\fHI=8.22458\ {\rm s}^{-1}$ is the two-photon rate for the transition $2s$-$1s$ according to \citet{Goldman}, $H(z)$ is the Hubble factor and $K_\fHI\equiv \lambda_{\fHI, 2p}^3/[8\pi H(z)]$ the cosmological redshifting of Lyman $\alpha$ photons.
\subsection{\HzI chemistry}
Due to the expansion of the universe, not all of the free electrons will recombine with protons. Instead, a freeze-out occurs, as recombination becomes less efficient at lower densities. The freeze-out abundance of free electrons was fitted by \citet{PeeblesPhys} as
\begin{equation}
\frac{n_{\fHII}}{n_\fHI}\sim1.2\times10^{-5}\frac{\Omega_0^{1/2}}{h\Omega_b}\label{elecfrac}
\end{equation}
for cosmologies with total mass density parameter $\Omega_0$, baryonic density parameter $\Omega_b$ and $h$ is the Hubble constant in units of $100\ \mathrm{km/s/Mpc}$. There is no contribution from helium to the free electron fraction, as helium recombines very early. The free electron fraction leads to the formation of molecules and ions like \HM and \HzII. As discussed by \citet{AnninosIn} and \citet{Anninos}, their abundances can be calculated by assuming chemical equilibrium, as their formation and destruction timescales are much shorter than the Hubble time. This yields the approximate expressions
\begin{eqnarray}
\frac{n_{\fHM}}{n_\fHI}&\sim& 2\times10^{-9}T^{0.88}\frac{n_{\fHII}}{n_\fHI},\\
\frac{n_{\fHzII}}{n_\fHI}&\sim&3\times10^{-14}T^{1.8}\frac{n_{\fHII}}{n_\fHI}.
\end{eqnarray}
The matter temperature can be calculated by assuming that at $z > 200$ it is the same as the radiation temperature, owing to efficient coupling of the temperatures through the Compton scattering of CMB photons off the residual free electrons, while at $z < 200$, where this coupling becomes ineffective, it evolves as for a simple adiabatic expansion \citep{Sunyaev, AnninosIn}. So
\begin{equation}
T = T_{0} (1+z),\ \  z>200, \quad T = \frac{T_0}{1+200}(1+z)^2,\ \ z<200,
\end{equation}
with $T_0=2.726$~K is the CMB temperature at $z=0$ \citep{fm02}. At redshifts $z\geq100$, 
$\fHM$ is efficiently photodissociated by the CMB and $\mathrm{H}_2$ is mainly formed by the process $\fHzII+\fHI\rightarrow \fHzI+\fHII$. Assuming that \HzII is formed most efficiently at redshift $z_0=300$ without being photo-dissociated by the CMB and that the hydrogen mass fraction is given by $f_\fHI=0.76$, one obtains for the \HzI abundance \citep{AnninosIn}
\begin{equation}
\frac{n_{\fHzI}}{n_\fHI}\sim2\times10^{-20}\frac{f_\fHI\Omega_0^{3/2}}{h\Omega_b}(1+z_0)^{5.1}.
\end{equation}
\subsection{Deuterium chemistry}
The \HDI abundance is mainly determined by the deuteration of hydrogen molecules (i. e. $\fHzI+\fDII\rightarrow\fHDI+\fHII$). It is thus crucial to have the correct abundance of \DIId, which is essentially determined by charge exchange with hydrogen atoms and ions, i.e.\ the processes $\fDI+\fHII\rightarrow\fDII+\fHI$ and $\fDII+\fHI\rightarrow\fDI+\fHII$.
As will be shown below, \DII is very close to chemical equilibrium, which yields the abundance
\begin{equation}
\frac{n_{\fDII}}{n_\fDI}\sim \mathrm1.2\times10^{-5}{\mathrm{exp}}\left(-43\ \mathrm{K}/T\right)\frac{\Omega_0^{1/2}}{h\Omega_b},\label{D+}
\end{equation}
when expression (\ref{elecfrac}) is used. The abundance of neutral deuterium atoms $n_\fDI$ can be determined by assuming that deuterium is almost fully neutral, i. e. $n_\fDI\sim\Omega_b f_\fDI \rho_c/m_\fDI$, where $f_\fDI$ is the total mass fraction of deuterium, $m_\fDI$ the mass of one deuterium atom, $\rho_c=3H_0^2/8\pi G$ the critical density and $G$ is Newton's constant. It is clear from expression (\ref{D+}) that the abundance of \DII drops exponentially at low temperatures. Direct deuteration of molecular hydrogen can thus only occur at redshifts where the exponential term is still of order one, before the exponential fall-off becomes significant. We thus evaluate the relative abundance at redshift 90. For both the formation and the destruction process $\fHDI+\fHII\rightarrow\fHzI+\fDII$, we estimate the rates with the simple expressions of \citet{Galli}. Again, we emphasize that more detailed numerical calculations should use the rates given in the appendix. The abundance is then given at $z=90$ as
\begin{eqnarray}
\left(\frac{n_{\fHDI}}{n_H}\right)_{z=90}&\sim&1.1\times10^{-7}f_D\mathrm{exp}(421\ \mathrm{K}/T_{z=90}) \frac{f_\fHI\Omega_0^{3/2}}{h\Omega_b} .
\end{eqnarray}
As there is no efficient destruction mechanism for \HDI at lower redshifts, the fractional abundance remains almost constant for $z<90$.
\subsection{\HeHII chemistry}
As we will show below in more detail, chemical equilibrium is also an excellent approximation to determine the abundance of \HeHIId. For the rates presented in \citet{Galli}, the process of stimulated radiative association of \HII and \HeI dominates over the non-stimulated rate. With the new rates presented in appendix \ref{heliumchem}, we find that both rates roughly coincide for gas temperatures greater than 10 K. Thus, for an analytic estimate, we approximate the combined formation rate through stimulated and non-stimulated radiative association by taking twice the rate for non-stimulated radiative association. The dominant destruction process at low redshifts is charge-exchange via $\fHeHII+\fHI\rightarrow\fHeI+\fHzII$, which yields for chemical equilibrium
\begin{eqnarray}
n_{\fHeHII}&\sim& 1.76\times10^{-10}n_{\fHeI}(n_{\fHII}/n_\fHI)\left(\frac{T}{300\ \mathrm{K}}\right)^{-0.24}\mathrm{exp}\left(-\frac{T}{4000\ \mathrm{K}}\right)\nonumber\\
&\sim& 7.03\times10^{-16}n_{\fHI}\frac{\Omega_0^{1/2}}{h\Omega_B}\left(\frac{T}{300\ \mathrm{K}}\right)^{-0.24}\mathrm{exp}\left(-\frac{T}{4000\ \mathrm{K}}\right).
\end{eqnarray}

\section{The chemical network}\label{Network}
In our chemical network, we have included the formation paths of \HzI and \HDId. The complete list of rates is given in table \ref{rates}, and some new rates that are relevant for \HeHII are discussed in more detail in the appendix. \HzI can be formed by two main channels, via the reactions $\fHM+\fHI\rightarrow \fHzI+\fe$ and $\fHzII+\fHI\rightarrow \fHzI+\fHII$. A very good compilation for the \HzI chemistry was given by \citet{Yoshida06}. Our compilation for the \HzI formation rates is similar, but we do not include all of their three-body-processes, as they are not relevant in the low-density regime explored here. Also, we keep those modifications for low temperatures that were originally given by \citet{Abel}. The ionized fraction of hydrogen is not determined by solving rate equations, but from the RECFAST code of \citet{SeagerFast}. The photodissociation rates for most of the hydrogen species are those of \citet{Galli}. For the photodissociation of \HzIId, we use the rate from their standard model, which assumes that the the levels of the molecule are populated according to LTE. Since the \HzIId level populations will be strongly coupled to the CMB at the redshifts at which \HzIId photodissociation is significant, this assumption is more reasonable than using a rate that assumes that \HzIId is completely in the ground state. However, a better understanding of this molecule would certainly be desirable. For the molecule \HDIId, we estimate the photodissociation due to reactions 22 and 23 of Table A.1 as half of the corresponding reaction for \HzIId. This is in agreement with recent isotopic helium experiments \citep{Pedersen, Buhr}, which found a similar effect for dissociative excitation with helium isotopes. %While the rates for the reactions
%\begin{eqnarray}
%\ ^3\mathrm{\fHeI}^4\mathrm{\fHeI}^+  +  \mathrm{\fe}  \rightarrow  \ ^3\mathrm{\fHeI} +\ ^4\mathrm{\fHeI}^+ +\mathrm{\fe},\\
%\ ^3\mathrm{\fHeI}^4\mathrm{\fHeI}^+  +  \mathrm{\fe}  \rightarrow  \ ^4\mathrm{\fHeI} +\ ^3\mathrm{\fHeI}^++\mathrm{\fe},
%\end{eqnarray}
%were roughly equal, the reaction rate for
%\begin{equation}
%\ ^4\mathrm{\fHeI}_2^+   + \mathrm{\fe}  \rightarrow  \ ^4\mathrm{\fHeI} +\ ^4\mathrm{\fHeI}^+ +\mathrm{\fe}
%\end{equation}
%was about twice as large as the single rates.
\\ \\ Photodissociation of \HzI by the Solomon process (reaction 20; Stecher and Williams, 1967) is calculated following the procedure described in \citet{Glover}, with the assumption that the rotational and vibrational levels of H2
   have their LTE level populations.
\\ \\
In a recent work of Capitelli et~al.\ \citep{Capitelli}, it was shown that the reaction rate for the process $\fHzI+\fe\rightarrow \fHM+\fHI$ is larger by several orders of magnitude than the rate given by \citet{Galli}. This is because they include the vibrational levels of molecular hydrogen in their calculation, and find that the excited vibrational levels cannot be neglected for this process. The change in the order of magnitude, however,  does not lead to a significant change in the results, as this rate is multiplied by the rather small densities of molecular hydrogen and free electrons.\\ \\
The hydrogen and helium chemistry is almost completely decoupled, as can be seen from the small rates for the charge exchange reactions 50 and 51. Their main interaction is via the \HeHII molecule. Owing to its relatively high contribution to the optical depth for the CMB, we have included this molecule in our chemical network and present some new rate coefficients for it in the appendix. The molecule \HDI gives a contribution to the cooling and is mainly formed through by the deuteration reactions 30 and 32, but there are also contributions from reactions 31, 43 and 44, involving \HDII, \DM and \HM. For the formation of deuterated molecules, it is therefore important to determine the ionized fraction of deuterium, which is given through the charge exchange reactions 27 and 28. Our deuterium network is inspired by the compilation of \citet{Nakamura}. The main difference compared to the deuterium network of \citet{Galli} is the detailed treatment of \DMd. We have added reaction 44, estimating its rate from reaction 43, thus also considering the contribution of \HMd. Also, reactions 27 and 28 are calculated from the revised rates of \citet{Savin}. As the fit provided by \citet{Savin} for the rate of reaction 27 becomes negative for $T<2.5\ \mathrm{K}$, it is set to zero at these temperatures\footnote{Note that this does not introduce a significant error, as owing to its exponential dependence on temperature, rate 27 is tiny at $T < 2.5 \: {\rm K}$.}. For temperatures larger than 200 K, the revised set of deuterium rates of Galli and Palla \citep{GalliDeut} are used. For lower temperatures, however, some of the new rates show unphysical divergences. In these cases, we use the rates from \citet{Galli} when the temperature drops below 200 K.

\section{The numerical algorithm}\label{Enzo}
To determine the evolution of primordial molecules in the early universe, we have employed the chemical network of the Enzo code \citep{O'SheaEnzo, Bryan, Norman} and extended the numerical approach developed by \citet{Anninos} for primordial chemistry to determine the chemical evolution of the homogeneous universe. The main issue was the calculation of the ionization fraction, which is determined by complex interactions between the CMB and the ground state as well as the excited states of atomic hydrogen, and which goes beyond typical applications of primordial chemistry. We thus included the RECFAST code of \citet{SeagerFast} as a subroutine for this calculation. For the deuterium and helium species as well as the molecules, however, we use the first order backwards differencing (BDF) method developed by \citet{Anninos}. The chemical timestep is set to $1\%$ of the hydrodynamical timestep. The latter is given by $\Delta t_{\rm hydro}=\eta(\Delta x / c_{\rm s})$, where $\Delta x$ is the cell size, $c_{\rm s}$ is the sound speed and $\eta$ is a safety factor, here taken to be 0.5. This proved sufficient to resolve the relevant chemical timescales, and simulations performed with even smaller timesteps gave identical results. The rate equations for the species $i$ are given in the form
\begin{equation}
\frac{dn_i}{dt}=-D_i n_i+C_i,\label{rate}
\end{equation}
where $D_i$ and $C_i$ are the destruction and creation coefficient for species $i$, respectively, which in general depend on the number densities of the other species and on the radiation field. Equation \ref{rate} is discretized and the right-hand-side is evaluated at the new timestep, yielding 
 \begin{equation}
 \frac{n^{\mathrm{new}}-n^{\mathrm{old}}}{\Delta t}=-D_i^{\mathrm{new}}n^{\mathrm{new}}+C_i^{\mathrm{new}}.
 \end{equation}
This can be solved for $n_i^{\mathrm{new}}$:
\begin{equation}
n_i^{\mathrm{new}}=\frac{C^{\mathrm{new}}\Delta t+n_i^{\mathrm{old}}}{1+D^{\mathrm{new}}\Delta t}. \label{noneq}
\end{equation}
The coefficients $D_i^{\mathrm{new}}$ and $C_i^{\mathrm{new}}$ are in general not known, but can be approximated using the species densities from the old timestep, and those species from the new timestep which have already been evaluated. \citet{Anninos} has argued that \HM and \HzII can even be evaluated assuming chemical equilibrium, since their reactions rates are much faster than those of the other species. Assuming some species $j$ in chemical equilibrium essentially means that $\dot{n}_j=0$, yielding
\begin{equation}
n_{j,eq}=\frac{C_j}{D_j}.\label{equi}
\end{equation}
In our chemical model, we have some additional species which have sufficiently fast reaction rates: \HeHIId, \DM and \HDIId. To check the validity of this assumption, we calculate the formation of molecules in two simulations: one using the non-equilibrium prescription (\ref{noneq}) for all species, and one using the equilibrium description (\ref{equi}) for the species with fast reaction rates. The results are presented in section \ref{Results} and confirm that the abundances of these species can be derived assuming chemical equilibrium. \\ \\

The BDF method is not a fully implicit numerical scheme, as several of the destruction and creation mechanisms must be approximated using the species densities from the previous timestep. However, we found that it is stable when the ionized fraction is provided from RECFAST. For consistency checks, we have varied the chemical timestep and explicitly ensured mass conservation for hydrogen, helium and deuterium, yielding consistent results. Note that mass conservation, charge conservation and positivity must be ensured as described by \citet{Anninos} if the ionized fraction is not provided from an independent routine. \\ \\We summarize the numerical algorithm in the following way:

\begin{itemize}
\item Loop over the chemical timestep $t_{chem}=0.01 t_{hydro}$ until the species have been evolved through the total hydrodynamical timestep.
\item Update the temperature and the abundances of \HId, \HII and \e with RECFAST.
\item Calculate the abundances of the atomic and ionized helium species using the non-equilibrium prescription (\ref{noneq}), and the abundance of the molecule \HeHII using (\ref{equi}).
\item Calculate the abundances of \HM and \HzII using (\ref{equi}), and the abundance of \HzI using (\ref{noneq}).
\item Calculate \DI using (\ref{noneq}), \DII, \DM and \HDII using (\ref{equi}), and finally \HDI using (\ref{noneq}).
\end{itemize}

\section{Results from the molecular network}\label{Results}
\begin{figure}
{\hspace{-0.25cm}}
\includegraphics[scale=0.7]{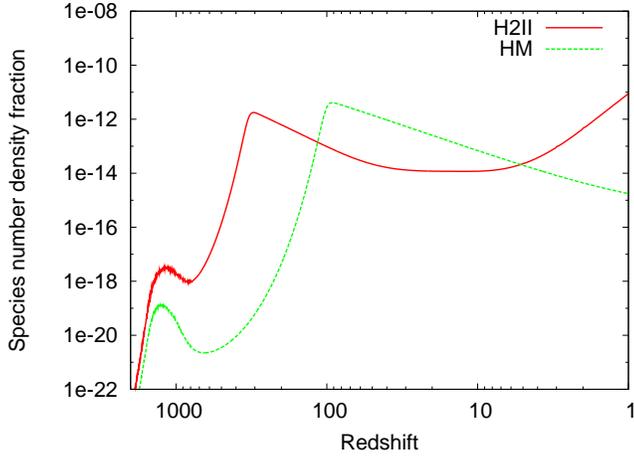}
\caption{Results for the evolution of \HzII and \HMd.}
\label{img:hydrocomp}
\end{figure}

\begin{figure}
{\hspace{-0.25cm}}
\includegraphics[scale=0.7]{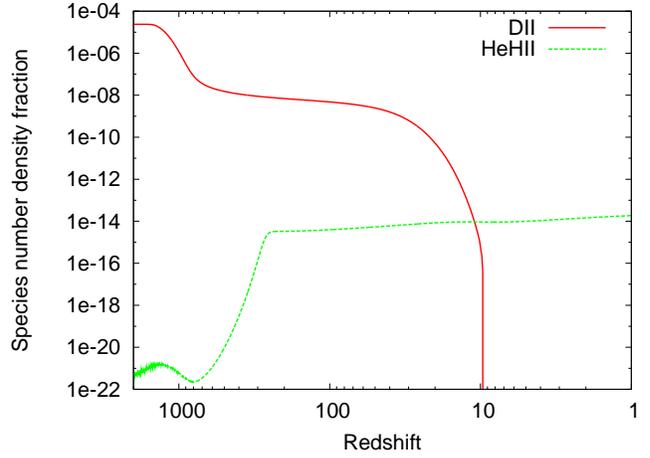}
\caption{Results for the evolution of \DII and \HeHIId.}
\label{img:DIIcomp}
\end{figure}

\begin{figure}
{\hspace{-0.25cm}}
\includegraphics[scale=0.7]{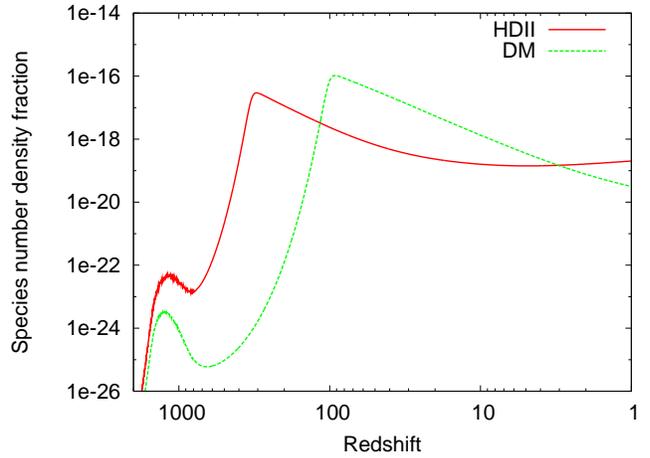}
\caption{Results for the evolution of \HDII and \DMd.}
\label{img:deutcomp}
\end{figure}

\begin{figure}
{\hspace{-0.25cm}}
\includegraphics[scale=0.7]{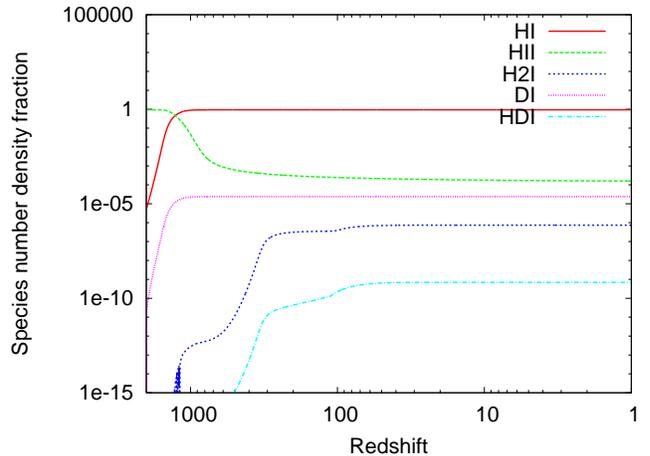}
\caption{Results for those species which freeze-out and are not in chemical equilibrium.}
\label{img:noneq}
\end{figure}

\begin{table}
    \begin{center}    
\begin{tabular}{|c|c|c|c|}
\hline
Cosmology / Ref. & [\e/\HI] & $[\fHzI/\fHI]$ & $[\mathrm{\fHDI/\fHI}]$  \\ \hline
WMAP 3 + SDSS & $2.63\times10^{-4}$  & $4.23\times10^{-7}$ & $2.39\times10^{-10}$  \\
\citep{Puynew}  & $\sim10^{-4}$  & $1.5\times10^{-7}$  & $\sim2\times10^{-10}$ \\
\citep{Hirata} & -          & $\sim2\times10^{-8}$  &     -           \\
%\citep{SeagerFast}     & $2.64\times10^{-4}$  & -  &  -  \\
Analytic approx.   & $2.0\times10^{-4}$  & $2.9\times10^{-7}$  & $1.6\times10^{-10}$  \\ 
%CDM  & $4.54\times10^{-4}$  & $2.14\times10^{-6}$  & $8.1\times10^{-10}$   \\ 
\citep{Galli} & $\sim2.4\times10^{-4}$  & $\sim1\times10^{-7}$ & $8\times10^{-11}$  \\ \hline
\end{tabular} 
\caption{Freeze-out of free electrons, \HzI and \HDI for different cosmologies at z=100. Results of other authors and the analytic approximations are included for comparison. WMAP3 + SDSS refer to the cosmological parameters  derived from the combined set of WMAP 3 + SDSS data \citep{Spergel}. \citep{Puynew, Hirata, SeagerFast} also refer to $\Lambda$CDM universes, while \citep{Galli} refer to a flat CDM universe.}
    \label{tab:freeze100}
    \end{center}
    \end{table}

  \begin{table}
    \begin{center}    
\begin{tabular}{|c|c|c|c|}
\hline
Cosmology / Ref. & [\e/\HI] & $[\fHzI/\fHI]$ & $[\fHDI/\fHI]$  \\ \hline
WMAP 3 + SDSS & $1.92\times10^{-4}$  & $7.97\times10^{-7}$ & $7.52\times10^{-10}$  \\
\citep{Puynew}  & $\sim6\times10^{-5}$  & $1.13\times10^{-6}$  & $3.67\times10^{-10}$ \\
\citep{Hirata} & -          &  $\sim6\times10^{-7}$ &     -           \\
\citep{SeagerFast}     & $1.92\times10^{-4}$  & -  &  -  \\
Analytic approx.  & $2.0\times10^{-4}$  & $2.9\times10^{-7}$  & $3.0\times10^{-10}$  \\ 
%CDM  & $4.54\times10^{-4}$  & $2.14\times10^{-6}$  & $8.1\times10^{-10}$   \\ 
\citep{Galli} & $3.02\times10^{-4}$ & $1.1\times10^{-6}$ & $1.21\times10^{-9}$ \\ \hline
\end{tabular} 
\caption{Freeze-out of free electrons, \HzI and \HDI for different cosmologies at z=10. Results of other authors and the analytic approximations are included for comparison. WMAP3 + SDSS refers to the cosmological parameters derived from the combined set of WMAP 3 + SDSS data \citep{Spergel}. \citep{Puynew, Hirata, SeagerFast} also refer to $\Lambda$CDM universes, while \citep{Galli} refer to a flat CDM universe.}
    \label{tab:freeze10}
    \end{center}
    \end{table}

\begin{table}
    \begin{center}    
\begin{tabular}{|c|c|c|}
\hline
Cosmology / Ref. & $[\fDII/\fHI]$ & $[\fHeHII/\fHI]$   \\ \hline
WMAP 3 + SDSS & $5.15\times10^{-9}$  & $4.27\times10^{-15}$  \\
\citep{Puynew}  & $\sim1.5\times10^{-9}$  &$\sim2.8\times10^{-15}$   \\
Analytic approx.      & $3.7\times10^{-9}$  & $4.5\times10^{-15}$   \\ 
%CDM  & $4.54\times10^{-4}$  & $2.14\times10^{-6}$   \\ 
\citep{Galli} & $\sim7.5\times10^{-9}$ & $\sim1.8\times10^{-14}$  \\ \hline
\end{tabular} 
\caption{Freeze-out of \DII and \HeHII for different cosmologies at z=100. Results of other authors  and our analytic approximations are included for comparison. WMAP3 + SDSS refers to the cosmological parameters derived from the combined set of WMAP 3 + SDSS data \citep{Spergel}. \citep{Puynew, Hirata, SeagerFast} also refer to $\Lambda$CDM universes, while \citep{Galli} refer to a flat CDM universe.}
    \label{tab:approx100}
    \end{center}
    \end{table}

\begin{table}
    \begin{center}    
\begin{tabular}{|c|c|c|}
\hline
Cosmology / Ref. & $[\fDII/\fHI]$ & $[\fHeHII/\fHI]$   \\ \hline
WMAP 3 + SDSS & $8.02\times10^{-16}$  & $9.98\times10^{-15}$  \\
\citep{Puynew}  & $\sim1\times10^{-19}$  & $4.6\times10^{-14}$  \\
Analytic approx.      & $2.1\times10^{-20}$  & $1.3\times10^{-14}$   \\ 
%CDM  & $4.54\times10^{-4}$  & $2.14\times10^{-6}$   \\ 
\citep{Galli} & $\sim5\times10^{-17}$  & $\sim6.5\times10^{-14}$  \\ \hline
\end{tabular} 
\caption{Freeze-out of \DII and \HeHII for different cosmologies at z=10. Results of other authors  and our analytic approximations are included for comparison. WMAP3 + SDSS refers to the cosmological parameters derived from the combined set of WMAP 3 + SDSS data \citep{Spergel}. \citep{Puynew, Hirata, SeagerFast} also refer to $\Lambda$CDM universes, while \citep{Galli} refer to a flat CDM universe.}
    \label{tab:approx10}
    \end{center}
    \end{table}

The formation of molecules after recombination was calculated for a $\Lambda$CDM model with $\Omega_{dm}=0.222$, $\Omega_b=0.044$, $\Omega_\Lambda=0.734$, $H_0=70.9\mbox{ }\mathrm{km/s/Mpc}$, $Y_p=0.242$, $[\fDI/\fHI]=2.4\times10^{-5}$, where $\Omega_{dm}, \Omega_b, \Omega_\Lambda$ are the density parameters for dark and baryonic matter as well as dark energy, $H_0$ is Hubble's constant, $Y_p$ is the mass fraction of helium with respect to the total baryonic mass and $[\fDI/\fHI]$ is the mass fraction of deuterium relative to hydrogen. These are the parameters from the combined set of WMAP 3-year data and the data of the Sloan Digital Sky Survey (SDSS) \citep{Spergel}. We expect only marginal changes in the results for the recently published WMAP 5-year data, as the cosmological parameters did not change significantly \citep{Komatsu08}. We have performed the calculation for the HD cooling functions of \citet{Galli}, \citet{Flower} and \citet{Lipovka}, and find that the results are not sensitive to this choice. For our ${\rm H_{2}}$ cooling function, we use that of \citet{Galli}, but  ${\rm H_{2}}$ cooling of the gas is never important, and our results
would not change significantly if we were to use the revised cooling rate of \citet{ga08}. The detailed evolution of the various species is plotted in Figs. \ref{img:hydrocomp}-\ref{img:noneq}. In Fig. \ref{img:hydrocomp}, we give the results for \HzII and \HMd, Fig. \ref{img:DIIcomp} shows the evolution of \DII and \HeHIId, and Fig. \ref{img:deutcomp} the evolution of \HDII and \DMd. For the other species, the results are given in Fig. \ref{img:noneq}. For several interesting species, we give the results at $z=100$ and $z=10$ in tables \ref{tab:freeze100}-\ref{tab:approx10}, and compare them to the results of \citet{Galli}, \citet{Puynew}, \citet{SeagerFast} and the analytical approximations of \citet{AnninosIn} and section \ref{Recombination}. As the analytic approximations do not take into account the \HM channel of \HzI formation, they underestimate the abundance found in the numerical calculation by roughly a factor of 2 at redshift 10.
\\ \\
The results clearly show the typical evolution of primordial chemistry as it is known from previous works. The two main channels for \HzI formation, via \HM and \HzIId, are reflected in its cosmic formation history, yielding a first major increase at $z\sim300$, where the relative abundance of \HzII reaches a maximum, and a second major increase at $z\sim100$, at a maximum of the \HM fraction. At redshifts below 5, there is a new rise in the abundance of \HzIId. This is likely an unphysical feature from the fit to the rate, which is not valid below 1 K. However, the real evolution at these low redshifts will in any case depart from our calculation due to reionization, metal enrichment and structure formation, and this feature is not relevant with respect to the CMB. The evolution of the deuterium species essentially follows the evolution of the hydrogen species. Since deuterium and hydrogen are strongly coupled via charge-exchange reactions, they recombine at almost the same time. However, due to the efficient charge-exchange reactions, there is no freeze-out of \DIId. Instead, its abundance drops exponentially at low redshifts. \DM and \HDII peak at the same redshifts as \HM and \HzII, and the evolution of \HDI resembles closely the evolution of \HzIId, as the dominant \HDI formation channel is given by $\fHzI+\fDII\rightarrow \fHDI+\fHII$. The evolution of \HeHII consists of a first phase where its evolution is determined by the effectiveness of photodissociation, and a second phase where it is determined by charge-exchange with neutral hydrogen atoms. Due to the new formation rates presented in section \ref{heliumchem}, formation through stimulated and non-stimulated radiative association of \HeI and \HII is almost equally important and in total less effective than found in previous works \citep{Puynew, Galli, Stancil}.
\\ \\
Numerically, the free electron fraction found in this work roughly agrees with the analytic approximations of \citet{AnninosIn} and previous results of \citet{Galli}, while \citet{Puynew} give a somewhat lower abundance. As the electrons act as catalysts for \HzI formation, this is reflected also in the abundance of molecular hydrogen, and similarly in the abundance of \HDI, as it is primarily formed by direct deuteration of molecular hydrogen. At $z=100$, the \HzI abundance found by \citet{Hirata} is still more than one order of magnitude below the abundance we find here, due to the effects of non-thermal photons, while at redshift 10, this effect is much less important and their result is only $25\%$ below the value found here. The analytic estimate of \citet{Anninos} also underestimates the \HzI abundance at $z=10$ by more than a factor of 2, as it does not take into account the \HzI formation via \HMd. At $z=100$, the abundances for \DII are of the same order of magnitude, although some differences exist, which may be due to the differences in the rates for the charge-exchange reactions, as well as differences in the abundance of ionized hydrogen. For the \HeHII molecule, the abundance is still comparable to previous results at redshift 100, but lower by a factor of 5 at $z=10$.\\ \\
We emphasize here that the abundance of \HM is similar to the one found by \citet{Black}. The peaks in the abundance at $z\sim100$ and $z\sim1400$ are reproduced, and their height agrees with the results from our calculation when the physical number density given in his paper is converted to the fractional abundance given in Fig. \ref{img:hydrocomp}. We estimate the uncertainty from reading off the values of his Fig. 1 to be a factor 2-3, and we point out that this is not sufficient to explain the difference in the optical depth that we find below.

\section{Effects of different species on the cosmic microwave background}\label{CMB}
\subsection{Molecular lines}
Due to the discreteness of the molecular lines, contributions to the optical depth arise only in narrow redshift intervals of the order $\Delta z/z\sim10^{-5}$, corresponding to the ratio between the thermal linewidth to the frequency of the transition. Peculiar motions of the order \textbf{$300\times(1+z)^{-1/2}\ \mathrm{km}\,\mathrm{s}^{-1}$} may further increase the effective linewidth by one or two orders of magnitude. This has no influence on the results presented here which depend only on the product of $\Delta z$ with the profile function that scales with the inverse of the linewidth, but may induce additional anisotropies, as we discuss in section \ref{discussion}. To evaluate Eq.~\ref{taumol} for a given frequency, we compute all the redshift intervals for which the photon frequency lies within an absorption line of the molecule, and add up the contributions from these frequencies. The relative importance of various molecules for the optical depth calculation can be estimated with the formulae of \citet{Dubrovich1994}. The most promising candidate is the \HeHII molecule, as it has a strong dipole moment and is formed from quite abundant species. Unfortunately, its destruction rate is very high as well. Another interesting molecule with a strong dipole moment is \HDIId. However, as a deuterated molecule, it has an even lower abundance. $\mathrm{H}_2\mathrm{D}^+$ is not considered because it has an even lower abundance \citep{Galli}, and $\mathrm{H}_3^+$ has both a low abundance and no dipole moment. The other molecule in our chemical network with a non-zero dipole moment is \HDId. However, its dipole moment is eight orders of magnitude smaller than that of \HeHIId, and so even though its peak abundance is five orders of magnitude larger, its effects will still be negligible compared to those of \HeHIId. We therefore do not consider it further. In spite of its strong dipole moment, $\mathrm{LiH}$ is also not considered, as it was already shown by \citet{Bougleux} and \citet{Galli} that its abundance is lower by roughly 10 orders of magnitude compared to the value assumed by \citet{Maoli94}, who discussed its potential relevance.\\ \\

For \HeHIId, we calculate the optical depth by using the large dataset provided by \citet{Engel}, which allows one to derive the line cross sections from the Einstein coefficients. The line cross section $\sigma_{M,i}$ weighted by the level population for a transition from an initial state $i$ with vibrational quantum number $v''$, rotational quantum number $J''$ and energy $E''$, to a final state $f$ with vibrational quantum number $v'$ and rotational quantum number $J'$ is given by\\
\begin{eqnarray}
\sigma_{M,i}(\nu)&=&\frac{1.3271\times10^{-12}(2J'+1)c^2}{Q_{vr}\nu^2}\mathrm{exp}\left(-\frac{-E''}{kT_r}\right)\nonumber\\
&\times&\left[1-\mathrm{exp}\left(-\frac{h_p\nu}{kT_r}\right) \right]A_{fi} \Phi(\nu-\nu_{fi}),\label{Engelsigma}
\end{eqnarray}
$A_{fi}$ is the Einstein coefficient, $\nu_{fi}$ is the frequency at line centre of the transition $i$, $\Phi(\nu-\nu_{fi})=\frac{1}{\Delta\nu_D\sqrt{\pi}}\mathrm{exp}(-(\nu-\nu_{fi})^2/\Delta\nu_D^2)$ the profile function for the line width $\Delta\nu_D=\sqrt{2kT/m_M}\nu_{fi}/c$, $m_M$ the mass of the molecule and $Q_{vr}$ is the partition function, given by $Q_{vr}=\sum_i g_i \mathrm{exp}(-E_i/kT)$ with the degeneracies $g_i=2J+1$. Again, we emphasize that the results are insensitive to the choice of the linewidth. The total cross section $\sigma_M$ is obtained as a sum over all $\sigma_{M,i}$. For temperatures between 500 and 10000 K, we use the fit provided by \citet{Engel}, while for lower temperatures, we do linearly interpolate between the values in their Table 5. The factor $\left[1-\mathrm{exp}\left(-h_p\nu/kT_r\right)\right]$ takes into account the correction for stimulated emission, which is especially relevant for the pure rotational transitions with low frequencies. As discussed in section \ref{effects}, this must be taken into account as stimulated emission does not change the direction of the emitted photons. \\ \\

 For \HDIId, we use the same formalism as for \HeHIId, but we determine the partition function from the accurate energy levels given by \citet{Karr}. Following \citet{Shu}, we use the transition moments $|D_{fi}|$ given by \citet{Colbourn} to determine the Einstein coefficients for the ro-vibrational lines as
\begin{equation}
A_{fi}=\frac{32\pi^3\nu_{fi}^3}{3\hbar_p c^3}|D_{fi}|^2,
\end{equation}
where $\hbar_p$ is the reduced Planck constant.

The Einstein coefficients for the pure rotational transitions are calculated with the dipole moment $D_0=0.86\ \mathrm{Debye}$ of \citet{Dubrovich1994} from
\begin{equation}
A_{0,J\rightarrow 0,J-1}=\frac{32\pi^3\nu_{fi}^3}{3\hbar_p c^3}D_0^2\frac{J}{2J+1}.
\end{equation}

\subsection{The negative hydrogen ion.}
There are two effects associated with the negative hydrogen ion than can affect the optical depth seen by CMB photons: the bound-free process of photodetachment that has also been discussed by \citet{Black}, and free-free transitions that involve an intermediate state of excited \HMd, i. e.
\begin{equation}
\fHI+\fe+\gamma\rightarrow \left(\fHM\right)^* \rightarrow \fHI+\fe. 
\end{equation}
While the importance of the free-free process is well-known for stellar atmospheres, there has been little work on this process in the low-temperature regime. As the fit formulae given by \citet{JohnH} and \citet{Gingerich} diverge at low temperatures, we have updated previous work of \citet{Dalgarno} to calculate the free-free absorption coefficient for the low temperature regime as described in appendix \ref{freefree}, and we use the fit of \citet{JohnH} to the calculation of \citet{Bell} for temperatures higher than $2000$ K, where it is accurate within $1\%$. For the bound-free process, we use the fit of \citet{JohnH} to the calculations of \citet{Wishart}. The treatment regarding absorption, spontaneous and stimulated emission is based on the expressions of \citet{Ruden}, but with the updated cross sections mentioned above. Apart from the usual optical depth due to absorption, we introduce also effective optical depths due to stimulated and spontaneous emission. The following expressions have to be evaluated for the redshift $z$, and the frequency dependence is suppressed for simplicity. Of course, we emphasize that the observed frequencies at $z=0$ must be related correctly to the physical frequencies at higher redshifts. As usual, the contribution to absorption is given as
\begin{equation}
d\tau_{bf,abs}=+n_{\fHM} \sigma_{bf}ds,
\end{equation}
where $ds$ is the cosmological line element, $\sigma_{bf}$ the cross section for the bound-free transition and $n_{\fHM}$ the number density of \HMd. We further introduce the Planck spectrum $B_{T}(\nu)$ of temperature $T$ and frequency $\nu$. The contribution to the effective optical depth from stimulated emission is then given as
\begin{equation}
d\tau_{bf,stim}=-(n_{\fHM})_{LTE}\sigma_{bf}e^{-h_p\nu/kT}ds,
\end{equation}
where the LTE abundance $(n_{\fHM})_{LTE}$ of \HM is given as
\begin{equation}
(n_{\fHM})_{LTE}=n_e n_H \frac{\lambda_e^3}{4}e^{h_p\nu_0/kT},
\end{equation}
where $\lambda_e=\frac{h_p}{\sqrt{2\pi m_e k T}}$ is the thermal de Broglie wavelength and $\nu_0=0.754\ \mathrm{eV}$ the binding energy of \HMd.
Note that the LTE abundance must be used here, as the processes of spontaneous and stimulated emission depend on the actual density of electrons and hydrogen atoms, and in general the correction for stimulated emission cannot be included as a factor of $(1-\mathrm{exp}(-h\nu/kT))$ unless the $\fHM$ ion has its LTE abundance \citep{Ruden}. The effective optical depth due to spontaneous emission is further given as
\begin{equation}
d\tau_{bf,spon}=-(n_{\fHM})_{LTE}\frac{2h_p\nu^3}{c^2 B_{T_r}(\nu)}\sigma_{bf}e^{-h_p\nu/kT}ds.
\end{equation}
For the free-free effect, stimulated emission is already included in the rate coefficients $a_\nu(T)$ given by \citet{JohnH}, which are normalized to the number density of neutral hydrogen atoms and the electron pressure. The effective contributions to the optical depth from absorption and emission are then given as
\begin{eqnarray}
d\tau_{ff,abs}&=&+n_H n_e kT a_\nu(T)ds,\\
d\tau_{ff,em}&=&-n_H n_e k T a_\nu(T)ds\frac{B_{T_g}(\nu)}{B_{T_r}(\nu)}.
\end{eqnarray}
The absorption by free-free transitions is thus proportional to a black-body spectrum for the radiation temperature, while the emission produces a spectrum determined by the gas temperature. In the following, we will refer further to the optical depth from absorption, which we define as
\begin{equation}
\tau_{abs}=\int \left(d\tau_{bf,abs}+d\tau_{ff,abs}\right),
\end{equation}
and the effective optical depth 
\begin{equation}
\tau_{eff}=\int\left(d\tau_{bf,abs}+d\tau_{bf,stim}+d\tau_{bf,spon}+d\tau_{ff,abs}+d\tau_{ff,em}\right),\label{efftau}
\end{equation}
which includes the corrections for emission. While the optical depth due to absorption, $\tau_{abs}$, is essentially responsible for photon scattering and a change in the power spectrum according to Eq.~(\ref{changepower}), the effective optical depth $\tau_{eff}$ leads to a net change in the radiation flux. The resulting change in the CMB temperature can be obtained by a linear expansion as
\begin{equation}
\Delta T=\tau_{eff}(\nu)\frac{B_{T_r}(\nu)}{(\partial B_{T_r}(\nu)/\partial T)_{T_r}}=f(\nu)\tau_{eff}(\nu) T_r,
\end{equation}
where we have introduced a frequency dependent correction factor $f(\nu)$, which can be evaluated to first order as 
\begin{equation}
f(\nu)=(1-\mathrm{exp}\left[-h_p\nu/kT\right])\frac{kT_r}{h\nu}.
\end{equation}

\subsection{The negative helium ion.}
\HeM is to some degree similar to the \HMd. However, only the free-free process contributes in practice, as any bound states autoionize on a timescale of the order of hundreds of microseconds \citep{Holoien, Brage}. Thus, we take into account only the free-free process, which can be treated in the same way as for \HMd. We approximate the corresponding free-free coefficient by a power law proportional to $\nu^{-2}$ and normalize with the data given by \citet{JohnHe}. Such a treatment should be sufficient up to frequencies of $1000$ GHz, and thus for the frequency range interesting for the Planck satellite.

\subsection{Photodissociation of \HeHIId.}
\begin{figure}
{\hspace{-0.25cm}}
\includegraphics[scale=0.7]{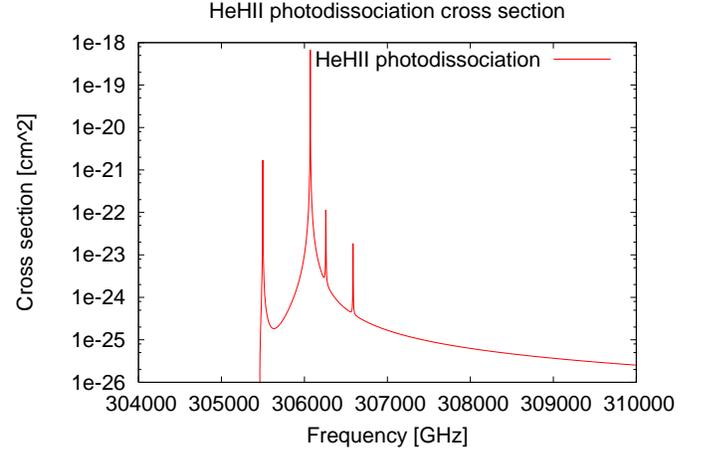}
\caption{\HeHII photodissociation cross section obtained from detailed balance.}
\label{img:HeHIIphoto}
\end{figure}
We used detailed balance to determine the photodissociation cross section from inverse reaction. The latter is essentially determined by several narrow resonances, which have been tabulated by Zygelman, Stancil and Dalgarno \citep{ZSD98}. The photodissociation cross section is thus given as the sum over resonances $i$ as
\begin{equation}
\sigma_{ph}(\nu)=\sum_i \frac{m_e c^2 (h_p\nu-E_0)}{(h_p\nu)^2}\frac{\Gamma_{r,i} \Gamma_i/2}{(h_p\nu-E_0-E_{r,i})^2+(\Gamma/2)^2},
\end{equation}
where the parameters $E_{r,i}$, $\Gamma_{r,i}$ and $\Gamma_i$ can be read off from Table 2 of \citet{ZSD98}, and $E_0$ is the photodissociation threshold for \HeHIId. From \citet{Dubrovich1997}, we adopt the value $E_0=1.85\ \mathrm{eV}$. The resulting photodissociation cross section is displayed in Fig. \ref{img:HeHIIphoto}. We consider only absorption, as it is sufficient to rule out the contribution from this molecule with respect to Planck.

\subsection{Observational relevance and results}\label{obs}
\begin{figure}
{\hspace{-0.25cm}}
\includegraphics[scale=0.7]{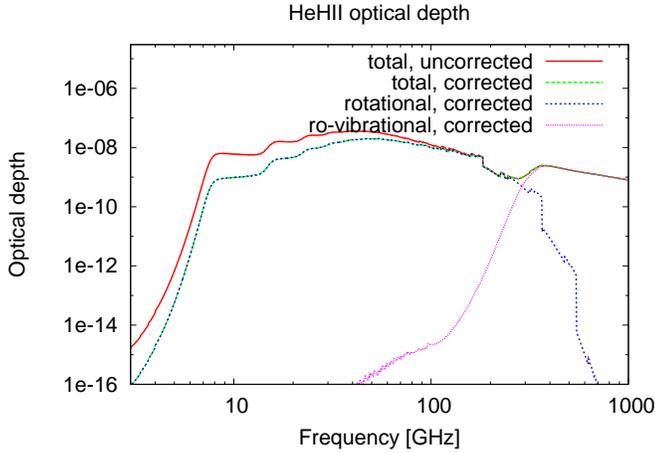}
\caption{The \HeHII optical depth, both corrected and uncorrected for stimulated emission. The contributions from pure rotational and ro-vibrational transitions are given separately. }
\label{img:tauhehpreal}
\end{figure}

\begin{figure}
{\hspace{-0.25cm}}
\includegraphics[scale=0.7]{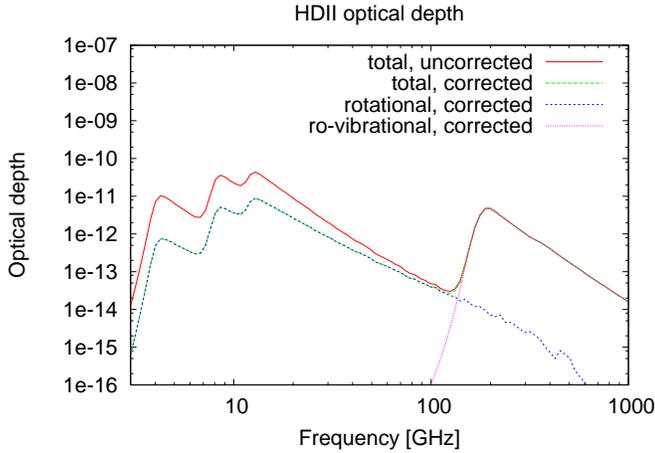}
\caption{The \HDII optical depth, both corrected and uncorrected for stimulated emission. The contributions from pure rotational and ro-vibrational transitions are given separately.}
\label{img:tauhdpreal}
\end{figure}
As discussed in section \ref{effects}, the optical depth from resonant scattering can effect the CMB by a change in the power spectrum according to Eq.~(\ref{changepower}), and it may also produce secondary anisotropies. We neglect the latter effect for the moment, as it is suppressed by the ratio of the peculiar velocity over the speed of light. To quantify the importance of \HeHII and \HDII for a change in the power spectrum, we have calculated the optical depth and corrected for stimulated emission as described in the previous subsection. The results are given in Figs. \ref{img:tauhehpreal} and \ref{img:tauhdpreal}. We find that the correction for stimulated emission is especially important for the lower frequencies of the pure rotational transitions. As explained by \citet{Basu04}, the sensitivity is not limited by cosmic variance when power spectra at different frequencies are compared, but the limit from instrumental noise corresponds to optical depths of $10^{-5}$ for the high-frequency bins. Thus, the signal is likely below the sensitivity of the Planck satellite by two orders of magnitude, but reasonable upper limits on the abundance of \HeHII are feasible. From Figs. B.1.a and B.2.a in the appendix, we estimate the uncertainty in the formation rate to be a factor of 2, while the destruction rate might be larger by up to an order of magnitude, if the old values of Roberge and Dalgarno \citep{RD82} are adopted. This defines the main uncertainty in this result. Even with an improved instrumental sensitivity, very accurate foreground subtraction would be required and may create additional noise. \\ \\

\begin{figure}
{\hspace{-0.25cm}}
\includegraphics[scale=0.7]{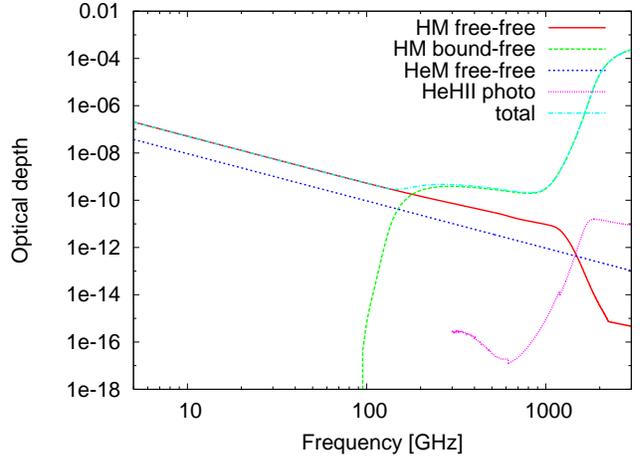}
\caption{The absorption optical depth due to different processes: the free-free processes of \HM and \HeMd, the bound-free process of \HMd, and the photodissociation of \HeHIId. Clearly, the total optical depth is dominated by the processes involving \HMd.}
\label{img:tauphoto}
\end{figure}

\begin{figure}
{\hspace{-0.25cm}}
\includegraphics[scale=0.7]{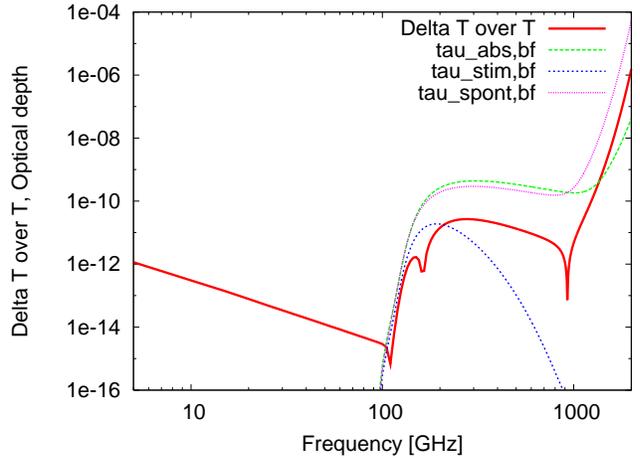}
\caption{The relative change in the CMB temperature due to the presence of \HMd. We further plot the optical depths due to absorption, spontaneous and stimulated emission for the bound-free process, as their overlap explains some features in the temperature change.}
\label{img:delta}
\end{figure}

%\begin{table}
%    \begin{center}    
%\begin{tabular}{|c|c|c|c|}
%\hline
%Frequency [GHz] & $\Delta \theta$ [arcmin] &  $\Delta T / T$   \\ \hline
%30 & 33.4   &$3.4\times10^{-14}$  \\
%44  & 26.8  & $1.5\times10^{-14}$ \\
%70  & 13.1  & $6.1\times10^{-15}$ \\ 
%100 & 9.2   & $2.9\times10^{-15}$\\ 
%143 & 7.1   & $1.5\times10^{-12}$\\ 
%217 & 5.0  & $1.9\times10^{-11}$\\ 
%353 & 5.0   & $2.3\times10^{-11}$\\ 
%545 & 5.0   & $1.1\times10^{-11}$\\ 
%857 & 5.0   & $2.6\times10^{-12}$\\ \hline
%\end{tabular} 
%\caption{\textbf{The expected relative change in the CMB temperature in the different frequency bins of Planck. The satellite can constrain deviations from a blackbody spectrum by comparison with a highly calibrated blackbody emitter in the low-frequency instrument and by coupling different frequency bins to the same cooling source. This should allow Planck to determine the absolute temperature at frequencies below 70 GHz with a precision of about $4\times10^{-7}$. See section \ref{obs} for more details.}}
%    \label{tab:planck}
%    \end{center}
%    \end{table}

Fig. \ref{img:tauphoto} shows the optical depth due to absorption by several free-free and photodestruction processes. We find that it is dominated at low frequencies by the free-free contributions of \HMd, and at high frequencies by the bound-free process of \HMd. The optical depth from the free-free processes of \HM and \HeM is essentially proportional to $\nu^{-2}$, at least for frequencies smaller than $1000$ GHz. This is essentially due to the characteristic frequency dependence in the absorption coefficient. Due to the approximation used for the \HeMd free-free absorption coefficient, the corresponding optical depth is somewhat overestimated for larger frequencies, but it is still only a subdominant contribution to the total optical depth. The effect from the bound-free transition dominates, but is significantly lower than previously reported by \citet{Black}, and the reason for this discrepancy is not obvious: as pointed out in section \ref{Results}, our \HM abundance agrees with his within a factor of 2 or 3, whereas the optical depth at high frequencies is different by three orders of magnitude. Although we have been unable to identify the reason for this disagreement, we have carefully checked our result and are confident that it is correct.\\ \\

As discussed in section \ref{effects}, photodestruction and free-free processes can change the net number of CMB photons. The calculated change in the CMB temperature due to the free-free and bound-free effect is given in Fig.~\ref{img:delta}, and depends on the effective optical depth defined in equation (\ref{efftau}). The change in the temperature due to the free-free transition is significantly lower than the absorption optical depth in Fig. \ref{img:tauphoto}, as the main contribution to the absorption optical depth comes from high redshifts $z>300$, where the difference between radiation and gas temperature is very small and the optical depth due to absorption and emission balance each other. For the bound-free transition, the absorption is balanced to some extend by the spontaneous emission, and for frequencies larger than $1000$ GHz, spontaneous emission in fact dominates, as the CMB flux at these frequencies is low by comparison. Near $150$ and $1000$ GHz, the contributions from spontaneous emission and absorption due to the bound-free process in fact become equal, making the net effect almost zero in a small frequency range. \\ \\

The net change in the CMB temperature due to the free-free and bound-free processes of \HM is small because absorption and emission processes are close to equilibrium. The optical depth due to absorption is thus considerably higher than the effective optical depth defined in Eq.~(\ref{efftau}), and can lead to a change in the power spectrum according to Eq.~(\ref{changepower}). Finally, we note that at redshifts where reaction 2 of Table B.1 dominates the destruction of $\fHM$, there is an uncertainty of up to an order of magnitude in our predicted $\fHM$ abundance, owing to the uncertainty in the rate of reaction 2 discussed in detail in \citet{GSJ06}. However, this uncertainty does not appear to significantly affect the size of our predicted signal, as the dominant contribution comes from redshifts at which reaction 2 is unimportant.

\section{Discussion and outlook}\label{discussion}
We have provided a detailed network for primordial chemistry and solved for the evolution in the homogeneous universe, and we examined the various ways in which primordial species can influence the CMB. The detailed calculation in the previous sections suggests that the \HM ion is only one order of magnitude below the detection threshold and strong upper limits on its abundance seem feasible, even though an accurate subtraction of frequency-dependent foregrounds will be required for this purpose. The relative deviations of the CMB from a pure blackbody have been constrained by \citet{Mather, Fixsen} to less than $1.5\times10^{-5}$. Distortions are also expected due to the two-photon process during recombination \citep{Chluba} and the helium and hydrogen lines \citep{Rubino2, Rubino1}. Finestructure transitions in heavy elements are also expect to produce some scattering after recombination \citep{Basu04}. An accurate measurement of distortions in the CMB and a precise measurement of the CMB power spectrum can thus improve our understanding of various processes during and after recombination if an accurate foreground subtraction is feasible. Given the tiny change in the CMB temperature found in this work, we further conclude that detecting the change in the power spectrum caused by scattering (see Eq.~(\ref{changepower})) is the most promising way to obtain constraints on the chemistry of the dark ages by future CMB experiments. 

So far, we have only taken into account effects arising from the homogeneous universe. One might argue that the molecular abundances could be very different if there were a protogalaxy at the redshift of resonance. In fact, the redshift intervals for which resonance occurs are very narrow, corresponding to some $100$ pc. This might be further increased by peculiar motions and local turbulence. \citet{NLA06} demonstrated that \HDI in cold collapsed clouds can lead to a strong local fluctuation of the order $10^{-5}$. Given the small volume fraction of such clouds, we neglect their impact in the present paper, although we will examine the effect of such inhomogeneous fluctuations in future work. \\ \\
\citet{Dubrovich1997} suggested luminescence as an additional effect that may amplify the signal and lead to a strong frequency dependence. This effect is well-known for stars in reflection nebulae. It occurs at redshifts $z=300-100$, where the rotational lines lie in the extreme Rayleigh-Jeans wing of the CMB and the first vibrational line near its maximum. Inelastic scattering in local velocity fields might thus provide much stronger frequency-dependent fluctuations. It is convenient to estimate the effect with the formula given by \citet{Dubrovich1997}:
\begin{equation}
\frac{\Delta T_r}{T_r}=\left( \frac{\Delta T_r}{T_r}\right)_0 \frac{f_m}{10^{-10}}\frac{V_p}{30\ \mathrm{km}/\mathrm{s}}\frac{\Omega_b}{0.1},
\end{equation}
where $f_m$ is the fractional abundance of the molecule, $V_p$ the peculiar velocity which can be estimated as $V_p=V_p(0)/\sqrt{1+z}$, $V_p(0)=600\ \mathrm{km}/\mathrm{s}$, and the quantity $\left(\Delta T_r/T_r\right)_0$ can be conveniently read off from Fig. 2 of \citet{Dubrovich1997}. For \HeHIId, it yields a maximum effect of roughly $\Delta T_r/T_r\sim10^{-11}$, and $\Delta T_r/T_r\sim10^{-12}$ for \HDIId , which are clearly below the sensitivity of the Planck satellite.\\ \\
As discussed by \citet{Launay}, \HzI is in general formed in excited states, but quickly decays into the ground state. Thus, \HzI formation produces additional photons that may lie within the CMB radiation and thus produce distortions to the blackbody radiation. Prelimary estimates based on the transition between the first excited vibrational state and the ground state indicate that the effect yields a relative change in the CMB temperature of the order $10^{-15}$, and is thus negligible. \\ \\
At the end of the dark ages, reionization will dramatically change the chemical evolution of the intergalactic gas, and produce large regions of ionized gas. In such regions, other formation channels for molecules could be relevant, like $\fHeII+\fHI\rightarrow\fHeHII+\gamma$. The reaction rate of this channel is four orders of magnitude larger than the radiative association of \HII and \HeId. On the other hand, the influence of destruction processes, such as dissociative recombination and photodissociation, will also be enhanced. The details of the evolution will also depend on the ratio of stellar to quasar sources and the details of the transition from Pop~III to Pop~II stars. A detailed analysis of this contribution is beyond the scope of this work, but clearly there is the possibility that this epoch could further increase the optical depths of \HeHIId and $\fHM$. Other contributions may arise from heavy elements at low redshift. \citet{Basu04}, \citet{Hernandez} and \citet{Basu07} suggested that one could use the change in the power spectrum of the CMB to constrain the chemical evolution of the low redshift universe. In fact, during the formation of metals and in early star forming regions, additional effects may occur that leave an interesting imprint in the CMB. As described by \citet{HernandezOI}, oxygen pumping may change the CMB temperature in metal enriched environments in a similar way as the Wouthuysen-Field effect that is well-known from $21$ cm studies \citep{Wouthuysen, Field}, and the inhomogeneous distribution of metallicity in bubble-like structures may influence the CMB power spectrum as described by \citet{HernandezOII}. \citet{HernandezCMB} further studied the effect of resonant scattering during reionization and recombination. In addition, star-forming regions may perturb the primordial signal through dust and molecular emission, especially CO \citep{Righi, RighiDust}. Such a potentially rich phenomenology will of course require a very careful analysis and a clear assessment of the different frequency dependences of various effects once the required sensitivity is reached. In the mean time, the increasing sensitivity in instruments like Planck \citep{Bersanelli}, the South Pole Telescope\footnote{http://pole.uchicago.edu/} \citep{Ruhl} and the Atacama Cosmology Telescope\footnote{http://www.physics.princeton.edu/act/} \citep{Fowler} will allow at least to set upper limits that may constrain theories involving the dark ages, reionization and recombination.

 %Still, we want to emphasize the possibility to search for signatures from the dark ages in the CMB, and we hope that future instruments will shed light on this interesting epoch of the universe. Constraining the high-redshift universe for signatures from the dark ages will benefit most from third generation experiments such as the South Pole Telescope \citep{Ruhl} and the Atacama Cosmology Telescope \citep{Fowler}. 

\acknowledgements
We thank Christoph Federrath, Antonella Maselli, Raffaella Schneider, Jamie O'Sullivan and Andreas Wolf for exciting discussions on this topic. We acknowledge the helpful support of Brian O'Shea and Pascal Paschos via the Enzo user list on topics involving primordial chemistry and the interaction with radiation. We further thank Jean-Philippe Karr for his advice on the transitions of the \HDII molecule, David Flower for helpful discussions on collisional excitation of \HeHII and Phillip Stancil for his advice on the photodissociation of \HeHIId. This research is partially supported by the Marie Curie Research Training network ``Constellation" (contract number MRTN-CT-2006-035890) and the Heidelberg Graduate School of Fundamental Physics (HGSFP). The HGSFP is funded by the Excellence Initiative of the German Government (grant number GSC 129/1). DS \ further thanks the LGFG, the DFG (under grant SFB 439) and the INAF-Osservatorio Astrofisico di Arcetri for financial support. His work was further supported by the the European Community - Research Infrastructure Action under the FP6 "Structuring the European Research Area" Program (HPC-EUROPA project RII3-CT-2003-506079). RSK acknowledges support from the DFG via grants KL1358/1 and SFB 439. We thank the anonymous referee for valuable comments and suggestions.

\appendix

\section{Free-free transitions involving \HM}\label{freefree}
\begin{figure}
{\hspace{-0.25cm}}
\includegraphics[scale=0.45]{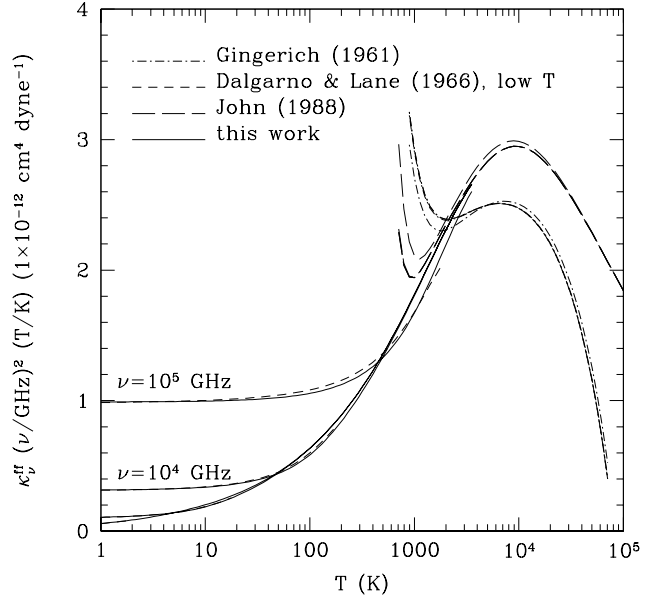}
\caption{The free-free absorption coefficient of \HM for $10^2$, $10^3$, $10^4$ and $10^5$ GHz, as a function of temperature. Given are the fits of \citet{JohnH} and \citet{Gingerich} for the high-temperature regime, the calculation of \citet{Dalgarno} based on effective range theory as well as the new calculation of this work for the low-temperature regime up to $2000$ K.}
\label{img:hmff}
\end{figure}
While the importance of free-free transitions is well-known in stellar atmospheres, this process is usually not considered in colder environments. Thus, high precision calculations are only available for stellar temperatures between $1000$ and $10000$ K \citep{JohnH, Bell, Gingerich}. The free-free absorption coefficient $k_{ff}$ is normalized per hydrogen atom and per unit electron pressure, and contributes to the optical depth by $d\tau=k_{ff}n_Hn_e kT ds$.  As the fits by \citet{JohnH} and \citet{Gingerich} diverge at low temperatures, thus giving an unphysical high contribution at low redshifts, we calculate the absorption coefficient based on the formalism of \citet{Dalgarno}, which takes into account only the contribution of the leading term to the transition moment. This approximation is valid at low energies and is thus a reasonable choice for the low-temperature regime. Assuming that the initial electron energies are described by a Maxwell distribution, the free-free coefficient $k_{ff}$ is given as
\begin{eqnarray}
k_{ff}(\nu)&=& 9.291\times 10^{-3} T^{-5/2} \left (1-e^{-a}\right) \int_0^{\infty}
E^{-1}\left(\frac{E}{h_p\nu}\right)^3\left(1+\frac{h_p\nu}{E}\right)^{1/2}\nonumber\\
&\times&e^{-aE/h_p\nu} \big(\left(1+\frac{h_p\nu}{E}\right)q_0(E)\nonumber\\
&+&q_0(E+h_p\nu) \big)dE\ \mathrm{cm}^4\ \mathrm{dyne}^{-1},
\end{eqnarray}
where $\nu$ is the photon frequency, $q_0(E)$ the zero-order elastic scattering cross section and $a=h_p\nu/kT$ a dimensionless parameter. While \citet{Dalgarno} used effective range theory \citep{Malley} to expand the cross section, we use the more accurate result of \citet{Dalgarno2} as fitted by \citet{Pinto}, given by
\begin{equation}
q_0(E)=\frac{4\times10^{-15}\ \mathrm{cm}^2}{\left(1+E/(3.8\mathrm{eV}) \right)^{1.84}}.
\end{equation}
In the range between $1$ and $10^7$ GHz and between $0.1$ and $2000$ K, our results are well-fit by the expression
\begin{equation}
k_{ff}(\nu)=10^{\left(a_1+a_2 x+a_3 x^2\right)\left(b_1+b_2 y+b_3 y^2\right)}\ \mathrm{cm}^4\ \mathrm{dyne}^{-1}\label{free-free},
\end{equation}
where $x=\log_{10}(\nu/\mathrm{GHz})$ and $y=\log_{10}(T/\mathrm{K})$, and $a_1=-3.2421$, $a_2=-0.502052$, $a_3=0.0117164$, $b_1=4.05293$, $b_2=0.169299$, $b_3=-0.00548517$. At stellar temperatures, our results differ by about $10\%$ from the calculation of \citet{Bell}, which is based on a more detailed treatment, whereas at lower temperatures, we expect an even higher accuracy of our result. We thus adopt the fit of \citet{JohnH} to the calculation of \citet{Bell} for $T>2000$ K and equation (\ref{free-free}) for lower temperatures.

%\longtabL{1}{
%\begin{landscape} 
%\begin{longtable}{|c|c|c|c|c|c|c|c|c|c|c|}
%\hline
%$\nu$ & $T=0.1$ & $T=0.32$ & $T=1.0$ & $T=3.2$ & $T=10$ & $T=32$ & $T=100$ & $T=320$ & $T=10^3$ & $T=3.2\cdot10^3$ \\ \hline
%1.0   & $2.03\cdot10^{13}$ & $1.15\cdot10^{13}$ & $6.45\cdot10^{12}$ & $3.63\cdot10^{12}$ & $2.04\cdot10^{12}$ & $1.14\cdot10^{12}$ & $6.37\cdot10^{11}$ & $3.49\cdot10^{11}$ & $1.81\cdot10^{11}$ & $8.22\cdot10^{10}$ \\ \hline
%\caption{The free-free absorption coefficient of the negative hydrogen ion, in units of $10^{-26}\ \mathrm{cm}^4\mathrm{dyne}^{-1}$. $\nu$ is the frequency in GHz, $T$ is the temperature in K.}
%\end{longtable}
%\end{landscape}
%}

\section{Reaction rates}\label{heliumchem}
Regarding the formation and destruction of \HeHII, we have updated some of the rate coefficients from the minimal model of \citet{Galli}.
\subsection*{Radiative association of \HeI and \HII}
The rate given by \citet{Galli} was a fit to the rate coefficients of Roberge and Dalgarno \citep{RD82} and Kimura et al. \citep{K93}. However, more accurate calculations are available from Ju{\v r}ek, {\v S}pirko and Kraemer \citep{JSK95} and Zygelman, Stancil and Dalgarno \citep{ZSD98}, which have been fitted by Stancil, Lepp and Dalgarno \citep{Stancil} for $10\ \mathrm{K}<T<10^4\ \mathrm{K}$, yielding
$8.0\times10^{-20}(T/300\ \mathrm{K})^{-0.24}\mathrm{exp}(-T/4000\ \mathrm{K})\ \mathrm{cm}^3\ \mathrm{s}^{-1}$, significantly below the rate of \citet{Galli} (see Fig. B.1.a).
\subsection*{Stimulated radiative association of \HeI and \HII}
The rate of stimulated radiative association depends on the background radiation, which is assumed to be a blackbody like the CMB. It thus depends both on the gas temperature $T$ and the radiation temperature $T_r$. It has been computed by \citet{JSK95} and \citet{ZSD98} and can be fitted for $10\ \mathrm{K}<T<10^4\ \mathrm{K}$ with the formula $3.2\times10^{-20}T^{1.8}(1.+0.1(T/1\ \mathrm{K})^{2.04})^{-1}\mathrm{exp}(-T/4000\ \mathrm{K})(1.+2\times10^{-4}(T_r/1\ \mathrm{K})^{1.1})\ \mathrm{cm}^3\ \mathrm{s}^{-1}$. The rate is plotted for various radiation temperatures in Fig. B.1.b.
\subsection*{Photodissociation of \HeHII into \HeI and \HII}
This rate can be determined from detailed balance of the reverse reaction (radiative association). Using the rate of \citet{JSK95} for radiative association yields a photodissociation rate slightly different from \citet{Galli}: $220(T_r/1\ \mathrm{K})^{0.9}\mathrm{exp}(-22740\ \mathrm{K}/T_r)\ \mathrm{s}^{-1}$ (c.f.\ Fig. B.2.a).
\subsection*{Proton transfer: $\fHeHII+\fHI\rightarrow \fHeI+\fHzII$}
\citet{Galli} adopted a constant rate for this reaction, which was based on a rate determination by Karpas, Anicich and Huntress \citep{KAH79} using the ion cyclotron resonance technique. A numerical integration of the new cross sections of \citet{LJB95} allows to give an improved rate with a slightly lower value, which can be fitted for $10^2\ \mathrm{K}<T<10^5\ \mathrm{K}$, yielding $0.69\times10^{-9}(T/300\ \mathrm{K})^{0.13}\mathrm{exp}(-T/33100\ \mathrm{K})\ \mathrm{cm}^3\ \mathrm{s}^{-1}$ (see Fig. B.2.b). \footnote{Note the discrepancy with \citet{Stancil}, who find a rate for the proton transfer reaction of \HeHII which is larger by a factor of 1.5.}
%\subsection*{Radiative association of \HI and \HeII}
%The rate given by \citet{Galli} was derived by fitting the rate coefficients of Zygelman and Dalgarno \citep{ZD90}. With the more recent coefficients of Kraemer, {\v S}pirko and Ju{\v r}ek \citep{KSJ95}, one obtains a slightly slower value, given by $1.6\times10^{-14}(T/1\ \mathrm{K})^{-0.33}\ \mathrm{cm}^3\ \mathrm{s}^{-1}$ (see Fig. A.3).

\begin{figure}
  \centering
   \subfigure[]{\includegraphics[scale=0.4]{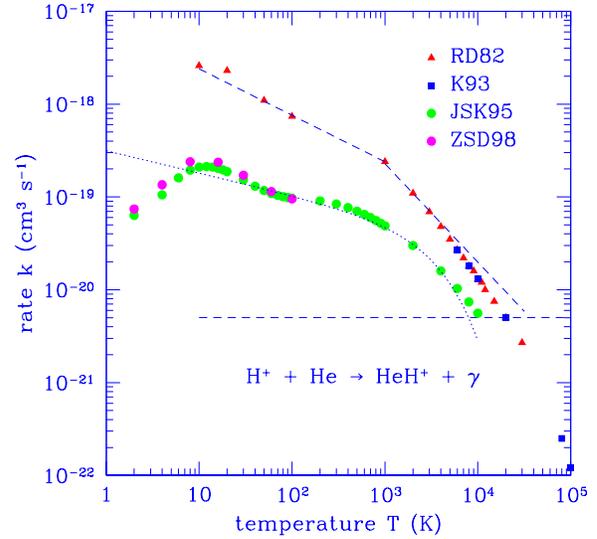}\label{he1}}
   \subfigure[]{\includegraphics[scale=0.4]{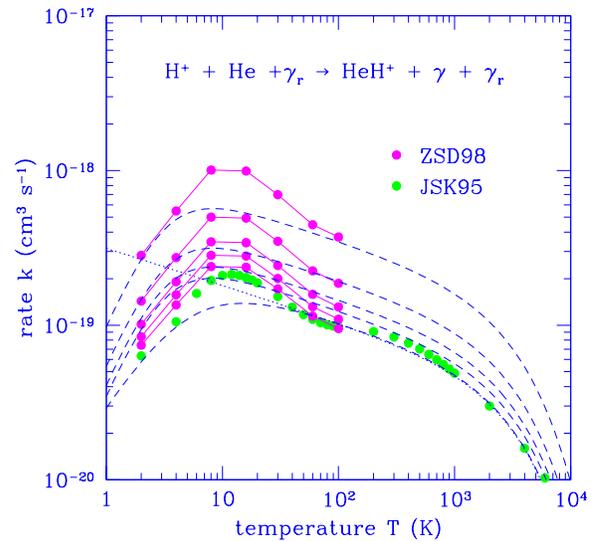}\label{he2}}
      \caption{New rate coefficients for helium chemistry. a) $\fHII+\fHeI\rightarrow \fHeHII+\gamma$ (rad. ass.). Dashed lines: Radiative association and inverse predissociation of \citet{Galli}, dotted line: \citet{Stancil}.
 b) $\fHII+\fHeI+\gamma\rightarrow \fHeHII+\gamma$. Stimulated association rate for black body radiation backgrounds with $T_r=0, 500, 1000, 2000$ and 5000 K.
} 
\end{figure}

\begin{figure}
  \centering
   \subfigure[]{\includegraphics[scale=0.4]{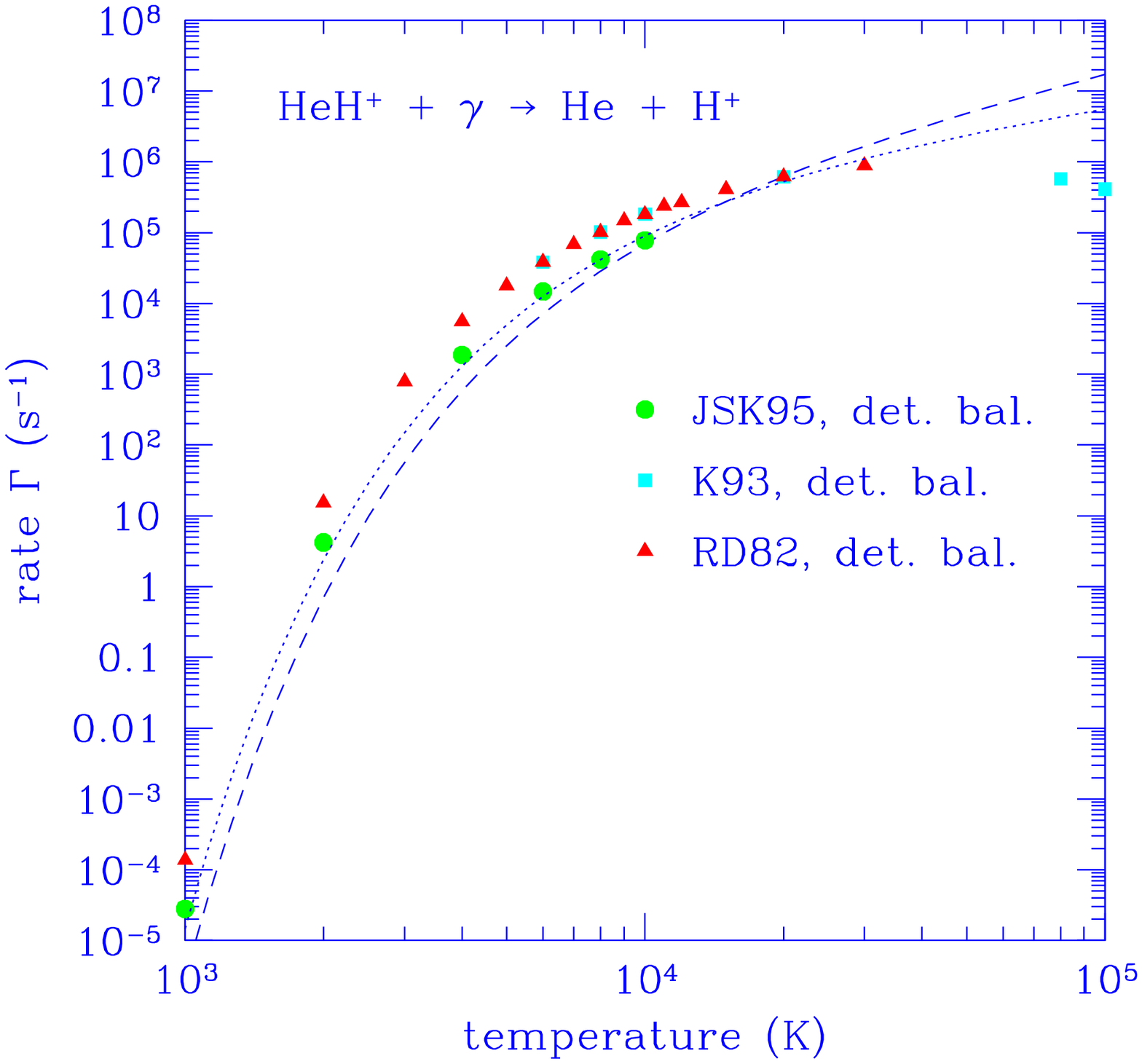}\label{he3}}
   \subfigure[]{\includegraphics[scale=0.4]{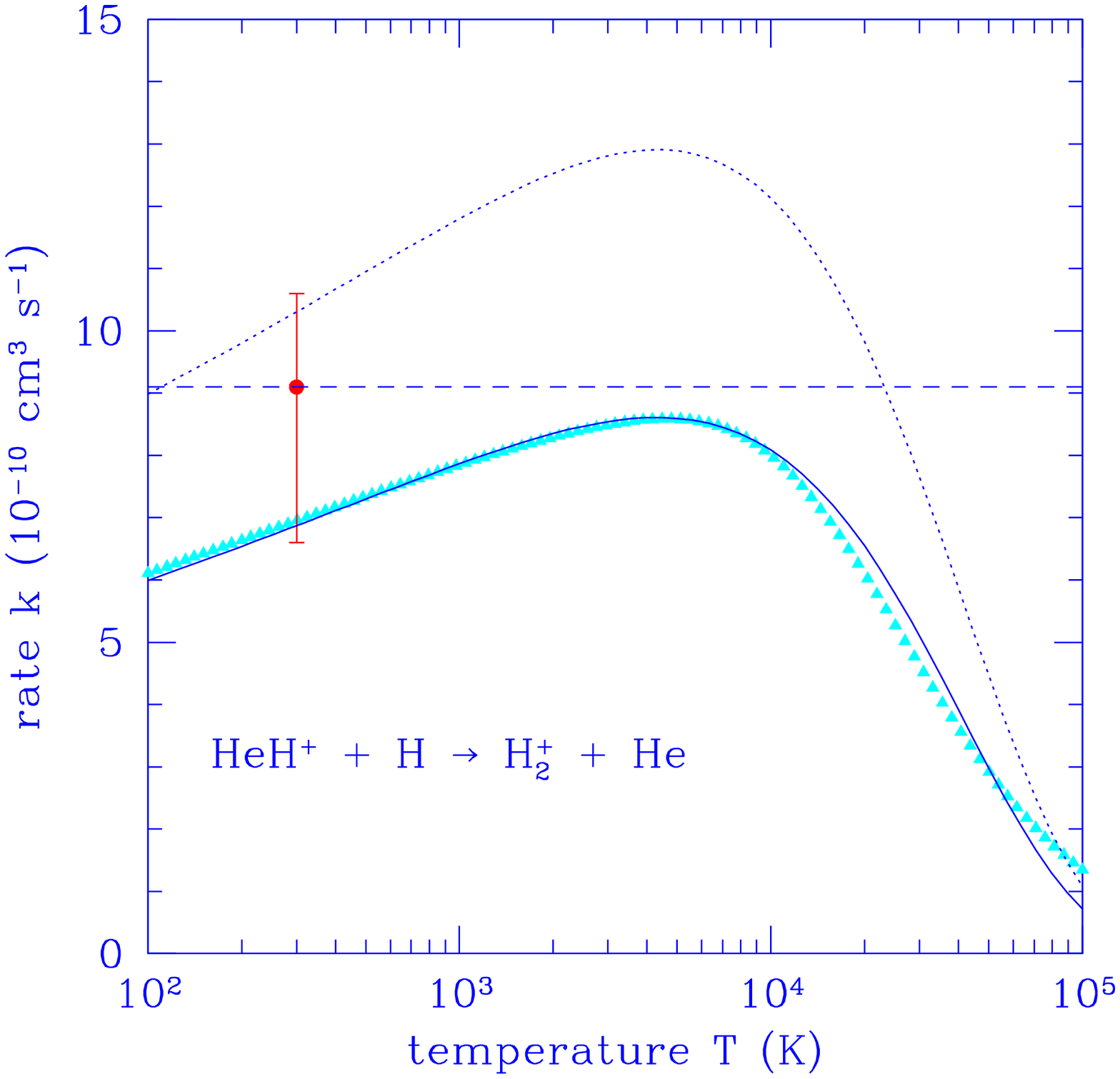}\label{he4}}\\
   \caption{New rate coefficients for helium chemistry. a) $\fHeHII+\gamma\rightarrow \fHeI+\fHII$. dashed line: \citet{Galli}, dotted line: \citet{Stancil}.
b) $\fHeHII+\fHI\rightarrow \fHeI+\fHzII$. Solid line: new fit, dotted line: \citet{Stancil}, dashed line: \citet{Galli}, dot: \citet{RD82}, triangles: \citet{LJB95} }
\end{figure}

%\begin{figure}
 % \centering

  % \subfigure[]{\includegraphics[scale=0.4]{heph_ra.ps}\label{he6}}
  %\caption{New rate coefficient for helium chemistry. $\fHeII+\fHI\rightarrow \fHeHII + \gamma$. Dashed line: \citet{Galli}, upper dotted line: new (recommended) fit, long-dashed line: fit to \citet{RD82}, lower dotted line: rate of \citet{Stancil}}
%\end{figure}

\longtabL{1}{
\begin{landscape}
\begin{longtable}{llll}
\caption{\label{rates} Collisional and radiative rates. T denotes the gas temperature in Kelvin, $T_{eV}$ the gas temperature in eV, $T_r$ the temperature of radiation in K. $\mathrm{dex}(x)=10^x$. Acronyms: AAZN97: Abel, Anninos, Zhang, Norman (1997), CCDL07: Capitelli, Coppola, Diomede, Longo (2007), GJ07: Glover and Jappsen (2007), GP98: Galli and Palla (1998), GP02: Galli and Palla (2002), JSK95: Ju{\v r}ek, {\v S}pirko, Kraemer (1995), LJB95: Linder, Janev, Botero (1995), LSD02: Lepp, Stancil, Dalgarno (2002), MSM96: Martin, Schwarz, Mandy (1996), PSS83: P\'{e}quignot, Petitjean, Boisson C. (1991), SKHS04b: Savin, Krstic, Haiman, Stancil (2004, Erratum), SLD98: Stancil, Lepp, Dalgarno (1998), TT02: Trevisan and Tennyson (2002), ZDKL89: Zygelman, Dalgarno, Kimura, Lane (1989), ZSD98: Zygelman, Stancil, Dalgarno (1998)}\\
\hline\hline
Reaction number & Reaction & Rate [cgs] & Reference \\ \hline
\endfirsthead
\caption{Collisional and radiative rates in cgs units. T denotes the gas temperature in Kelvin, $T_{eV}$ the gas temperature in eV, $T_r$ the temperature of radiation in K. $\mathrm{dex}(x)=10^x$. Acronyms: AAZN97: Abel, Anninos, Zhang, Norman (1997), CCDL07: Capitelli, Coppola, Diomede, Longo (2007), GJ07: Glover and Jappsen (2007), GP98: Galli and Palla (1998), GP02: Galli and Palla (2002), JSK95: Ju{\v r}ek, {\v S}pirko, Kraemer (1995), LJB95: Linder, Janev, Botero (1995), LSD02: Lepp, Stancil, Dalgarno (2002), MSM96: Martin, Schwarz, Mandy (1996), PSS83: P\'{e}quignot, Petitjean, Boisson C. (1991), SKHS04b: Savin, Krstic, Haiman, Stancil (2004, Erratum), SLD98: Stancil, Lepp, Dalgarno (1998), TT02: Trevisan and Tennyson (2002), ZDKL89: Zygelman, Dalgarno, Kimura, Lane (1989), ZSD98: Zygelman, Stancil, Dalgarno (1998)}\\
\hline\hline
Reaction number & Reaction & Rate [cgs] & Reference \\ \hline
\endhead
\hline
\endfoot
 1 & \HId+\e$\rightarrow$ \HMd+$\gamma$ & \textbf{$1.4\times10^{-18}T^{0.928}\mathrm{exp}(-T/16200)$} & \cite{Galli}  \\ 
 %10 & $HM+HI \rightarrow H2I+e$ &    \textbf{   $exp(-20.06913897587003
 %               + 0.2289800603272916*\ln T_{eV}
 %               + 0.03599837721023835*(\ln T_{eV})^2
  %              - 0.004555120027032095*(\ln T_{eV})^3
   %             - 0.0003105115447124016*(\ln T_{eV})^4
    %            + 0.0001073294010367247*(\ln T_{eV})^5
     %           - 8.36671960467864d-6*(\ln T_{eV})^6
      %          + 2.238306228891639d-7*(\ln T_{eV})^7)$ for $T_{eV}>0.1$}, &  \\ 
 %&  &    \textbf{  $1.43\times10^{-9}$ for $T_{eV}<0.1 $} &   \cite{Abel} \\		
 2 & \HMd+\HI $\rightarrow$ \HzId+\e &    \textbf{   $1.5\times10^{-9}$} for \textbf{$T<300$} & \cite{Galli} \\ 
 &  &    \textbf{  $4.0\times10^{-9}T^{-0.17}$} for \textbf{$T>300 $} &   \cite{Galli} \\		
 
 %11 & $HI+HII\rightarrow H2II+\gamma$ & \textbf{$1.85\times10^{-23}*T^{1.8}$ } for $T<6700,$ & \cite{Abel}  \\ 
 3 & \HId+\HII$\rightarrow$ \HzII+$\gamma$ & \textbf{$\mathrm{dex}(-19.38-1.523\log_{10}(T)+1.118\log_{10}^2(T)-0.1269\log_{10}^3(T))$ }  & \cite{Galli}  \\ 
  
 %12 & $H2II+HI\rightarrow H2I+HII$ & \textbf{$6.0\times10^{-10}$} & \cite{Abel}  \\ 
 4 & \HzIId+\HI$\rightarrow$ \HzId+\HII & \textbf{$6.4\times10^{-10}$} & \cite{Galli}  \\ 
 %13 & $2HI+HI\rightarrow H2I+HI$ & \textbf{$1.3\times10^{-32} * (T/300.0)^{-0.38}$ for $T<300$}  & \cite{Abel} \\ 
  %&  & \textbf{$1.3\times10^{-32} * (T/300.0)^{-1}$ for $T>300$}  & \cite{Abel} \\ 
  
  5 & 2\HId+\HI$\rightarrow$ \HzId+\HI & \textbf{$5\times10^{-29}T^{-1}$} for \textbf{$T<300$}  & \cite{Palla} \\ 
  
 %12 & $H2I+HII\rightarrow H2II+HI$ & \textbf{$ exp(-24.24914687731536
 %                + 3.400824447095291*\ln T_{eV}
 %                - 3.898003964650152*(\ln T_{eV})^2
 %                + 2.045587822403071*(\ln T_{eV})^3
 %                - 0.5416182856220388*(\ln T_{eV})^4
 %                + 0.0841077503763412*(\ln T_{eV})^5
 %                - 0.007879026154483455*(\ln T_{eV})^6
 %                + 0.0004138398421504563*(\ln T_{eV})^7
 %                - 9.36345888928611d-6*(\ln T_{eV})^8)$} & \cite{Abel}  \\ 
6 & \HzId+\HII$\rightarrow$ \HzIId+\HI & %\textbf{$\mathrm{exp}(-21237.1/T)[-3.3232183\times10^{-7}+3..3735382\times10^{-7}\ln T-1.4491368\times10^{-7}\ln^2(T)
%+3.4172805\times10^{-8}\ln^3(T)-4.7813720\times10^{-9}\ln^4(T)+3.9731542\times10^{-10}\ln^5(T)-1.8171411\times10^{-11}\ln^6(T)+3.5311932\times10^{-13}\ln^7(T)]$} 
see reference & \cite{Savinb} \\
 %15 & $H2I+e\rightarrow 2HI+e$ &    \textbf{   $4.38\times10^{-10}*exp(-102000.0/T)*T^{0.35}$} & \cite{Abel} \\ 
 7 & \HzId+\e$\rightarrow$ 2\HId+\e &    \textbf{   $1.91\times10^{-9}T^{0.136}\mathrm{exp}(-53407.1/T)$} & \cite{Trevisan} \\ 
8 & \HzId+\HI $\rightarrow$ 3\HI &     Fit to data of \cite{Martin} &   \cite{Martin} \\		
9 & \HMd+\e$\rightarrow$ \HI+2\e &% \textbf{$         \mathrm{exp}(-18.01849334273
                 %+ 2.360852208681*\ln T_{\mathrm{eV}}
%                 - 0.2827443061704*(\ln T_{\mathrm{eV}})^2
 %                + 0.01623316639567*(\ln T_{\mathrm{eV}})^3
  %               - 0.03365012031362999*(\ln T_{\mathrm{eV}})^4
   %              + 0.01178329782711*(\ln T_{\mathrm{eV}})^5
    %             - 0.001656194699504*(\ln T_{\mathrm{eV}})^6
     %            + 0.0001068275202678*(\ln T_{\mathrm{eV}})^7
      %           - 2.631285809207d-6*(\ln T_{\mathrm{eV}})^8)$ }
see reference & \cite{Abel}  \\ 
 10 & \HMd+\HI$\rightarrow$ 2\HId+\e & %\textbf{$ 2.56\times10^{-9}*T_{\mathrm{eV}}^{1.78186} $ for $T_{\mathrm{eV}}<0.1$ } & \cite{Abel}  \\ 
% & & \textbf{$  \mathrm{exp}(-20.37260896533324
%                 + 1.139449335841631*\ln T_{\mathrm{eV}}
 %                - 0.1421013521554148*(\ln T_{\mathrm{eV}})^2
  %               + 0.00846445538663*(\ln T_{\mathrm{eV}})^3
   %              - 0.0014327641212992*(\ln T_{\mathrm{eV}})^4
    %             + 0.0002012250284791*(\ln T_{\mathrm{eV}})^5
     %            + 0.0000866396324309*(\ln T_{\mathrm{eV}})^6
      %           - 0.00002585009680264*(\ln T_{\mathrm{eV}})^7
       %          + 2.4555011970392d-6*(\ln T_{\mathrm{eV}})^8
        %         - 8.06838246118d-8*(\ln T_{\mathrm{eV}})^9)$ for $T_{\mathrm{eV}}>0.1$ } 
see reference& \cite{Abel}  \\ 		 
 %19 & $HM+HII\rightarrow 2HI$ & \textbf{$6.5\times10^{-9}/\sqrt{T_{eV}}$} & \cite{Abel}  \\ 
 11 & \HMd+\HII$\rightarrow$ 2\HI & \textbf{$1.40\times10^{-7}(T/300)^{-0.487}\mathrm{exp}(T/29300)$} & \cite{Lepp2002}  \\ 
 %20 & $HM+HII\rightarrow H2II+e$ & \textbf{$1.0\times10^{-8}*T^{-0.4}$ for $T<10000$}  & \cite{Abel} \\ 
  %&  & \textbf{$4.0\times10^{-4}*T^{-1.4}*exp(-15100.0/T)$ for $T>10000$}  & \cite{Abel} \\ 
  12 & \HMd+\HII$\rightarrow$ \HzIId+\e & \textbf{$6.9\times10^{-9}T^{-0.35}$} for \textbf{$T<8000$}  & \cite{Galli} \\ 
     &                            & \textbf{$9.6\times10^{-7}T^{-0.9}$} for \textbf{$T>8000$}   & \cite{Galli} \\
 %21 & $H2II+e\rightarrow 2HI$ & \textbf{$5.56396\times10^{-8}/T_{eV}^{0.6035}$}  & \cite{Abel} \\ 
 13 & \HzIId+\e$\rightarrow$ 2\HI & \textbf{$2.0\times10^{-7}T^{-0.5}$}  & \cite{Galli} \\ 
 14 & \HzIId+\HM $\rightarrow$ \HI+\HzI & \textbf{$5\times10^{-6} T^{-0.5}$} for \textbf{$T>100$}   & \cite{Abel} \\
 15 & \HzId+\e$\rightarrow$ \HId+\HM & \textbf{$3.67\times10^{1}T^{-2.28}\mathrm{exp}(-\frac{47172}{T})$} & \cite{Capitelli} \\ 
 16 & \HMd+$\gamma$ $\rightarrow$ \HId+\e     &  $1.1\times10^{-1}T_r^{2.13}\mathrm{exp}(-8823/T_r)$  & \cite{Galli}    \\
 17 & \HzIId+$\gamma$ $\rightarrow$ \HId+\HII     &  $1.63\times10^7\mathrm{exp}(-32400/T_r)$  &  \cite{Galli}   \\
 18 & \HzId+$\gamma$ $\rightarrow$ \HzIId+\e     &  $2.9\times10^2T_r^{1.56}\mathrm{exp}(-178500/T_r)$  & \cite{Galli}    \\
 19 & \HzIId+$\gamma$ $\rightarrow$ 2\HIId+\e     &  $9.0\times10^1T_r^{1.48}\mathrm{exp}(-335000/T_r)$  &  \cite{Galli}   \\
 20 & \HzId+$\gamma$ $\rightarrow$ $\left(\fHzI\right)^*\rightarrow 2\fHI$ & $1.13\times10^6 T_r^{0.369}\mathrm{exp}(-140000/T_r)$ & \cite{Glover} \\
 21 & \DMd+$\gamma$ $\rightarrow$ \DId+\e      &  estimated by rate 16            &                \\
 22 & \HDIId+$\gamma$ $\rightarrow$ \DId+\HII   &  estimated by half of rate 17         &         \\
 23 & \HDIId+$\gamma$ $\rightarrow$ \HId+\DII    & estimated by half of rate 17        &     \\
 24 & \HDIId+$\gamma$ $\rightarrow$ \HIId+\DIId+\e  & estimated by rate 19     &       \\
  25 & \HDId+$\gamma$ $\rightarrow$ \HDIId+\e  & estimated by rate 18     &       \\
26  & \DIId+\e$\rightarrow$ \DI+$\gamma$ & \textbf{$3.6\times10^{-12}(T/300)^{-0.75}$} & \cite{Stancil}     \\
27  & \DId+\HII$\rightarrow$ \DIId+\HI  & \textbf{$2\times10^{-10}T^{0.402}\mathrm{exp}(-37.1/T)-3.31\times10^{-17}T^{1.48}$} & \cite{Savin} \\
28  & \DIId+\HI$\rightarrow$ \DId+\HII & \textbf{$2.06\times10^{-10}T^{0.396}\mathrm{exp}(-33.0/T)+2.03\times10^{-9}T^{-0.332}$} & \cite{Savin}\\
29  & \DId+\HI$\rightarrow$ \HDId+$\gamma$ & see reference & \cite{Dickinson,DickinsonErr}\\
30  & \DId+\HzI$\rightarrow$ \HDId+\HI & \textbf{$1.69\times10^{-10}\mathrm{exp}(-4680/T+198800/T^2)$} for \textbf{$T>200$} & \cite{GalliDeut}\\
    &                            & \textbf{$9.0\times10^{-11}\mathrm{exp}(-3876/T)$} for \textbf{$T<200$} & \cite{Galli} \\
31  & \HDIId +\HI$\rightarrow$ \HDId+\HII & \textbf{$6.4\times10^{-10}$} & \cite{Stancil}\\
32  & \DIId+\HzI$\rightarrow$ \HDId+\HII & \textbf{$1.0\times10^{-9}(0.417+0.846\log_{10}(T)-0.137\log_{10}^2(T))$} & \cite{GalliDeut}\\ 
33  & \HDId+\HI$\rightarrow$ \DId+\HzI &\textbf{$5.25\times10^{-11}\mathrm{exp}(-4430/T+173900/T^2)$} for \textbf{$T>200$} & \cite{GalliDeut}\\
    &                            &\textbf{$3.2\times10^{-11}\mathrm{exp}(-3624/T)$} for \textbf{$T<200$}    & \cite{Galli}\\
34  & \HDId+\HII$\rightarrow$ \DIId+\HzI & \textbf{$1.1\times10^{-9}\mathrm{exp}(-488/T)$} & \cite{GalliDeut}\\
35  & \DId+\HII$\rightarrow$ \HDIId+$\gamma$ &\textbf{$\mathrm{dex}(-19.38-1.523\log_{10}(T)+1.118\log_{10}^2(T)-0.1269\log_{10}^3(T))$} & \cite{Galli}\\
36  & \DIId+\HI$\rightarrow$ \HDIId+$\gamma$  &\textbf{$\mathrm{dex}(-19.38-1.523\log_{10}(T)+1.118\log_{10}^2(T)-0.1269\log_{10}^3(T))$} & \cite{Galli} \\ 
37  & \HDIId+\e$\rightarrow$ \DId+\HI & \textbf{$7.2\times10^{-8}T^{-0.5}$}& \cite{Stancil}\\
38  & \DId+\e$\rightarrow$ \DM + $\gamma$ & \textbf{$3.0\times10^{-16}(T/300)^{0.95}\mathrm{exp}(-T/9320)$} & \cite{Stancil}\\
39  & \DIId+\DM$\rightarrow$ 2\DI & \textbf{$1.96\times10^{-7}(T/300)^{-0.487}\mathrm{exp}(T/29300)$} & \cite{Lepp2002}\\
40  & \HIId+\DM $\rightarrow$ \DId+\HI & \textbf{$1.61\times10^{-7}(T/300)^{-0.487}\mathrm{exp}(T/29300)$} & \cite{Lepp2002}\\
41  & \HMd+\DI$\rightarrow$ \HId+\DM & \textbf{$6.4\times10^{-9}(T/300)^{0.41}$} & \cite{Stancil}\\
42  & \DMd+\HI$\rightarrow$ \DId+\HM & \textbf{$6.4\times10^{-9}(T/300)^{0.41}$} & \cite{Stancil}\\
43  & \DMd+\HI$\rightarrow$ \HDId+\e & \textbf{$1.5\times10^{-9}(T/300)^{-0.1}$} & \cite{Stancil}\\
44  & \DId+\HM$\rightarrow$ \HDId+\e &  estimated by rate 42  & This work\\ 
45  & \HMd+\DII$\rightarrow$ \DId+\HI &\textbf{$1.61\times10^{-7}(T/300)^{-0.487}\mathrm{exp}(T/29300)$} & \cite{Lepp2002}\\
46  & \HeIIId+\e$\rightarrow$ \HeIId+$\gamma$ & \textbf{$3.36\times10^{-10}T^{-0.5}(T/1000)^{-0.2}\left(1+(T/10^6)^{0.7} \right)^{-1}$} &\cite{Cen} \\
47  & \HeIId+$\gamma$ $\rightarrow$ \HeIIId + \e & see reference &\cite{Abel} \\
48  & \HeIId+\e$\rightarrow$\HeId+$\gamma$ &see reference & \cite{Abel} \\
49  & \HeId+$\gamma$ $\rightarrow$ \HeIId+\e &see reference & \cite{Abel}\\
50  & \HeId+\HII $\rightarrow$ \HeIId+\HI & \textbf{$4.0\times10^{-37}T^{4.74}$} for \textbf{$T>10000$} & \cite{Galli}\\
    &              & \textbf{$1.26\times10^{-9}T^{-0.75}\mathrm{exp}(-127500/T)$} for \textbf{$T<10000$} & \citet{Glover}  \\
51  & \HeIId+\HI $\rightarrow$ \HeId+\HII & \textbf{$1.25\times10^{-15}\times\left(T/300 \right)^{0.25}$} & \cite{Zygelman}\\
52  & \HeId+\HII $\rightarrow$ \HeHIId+$\gamma$, radiative association & \textbf{$8.0\times10^{-20}(T/300)^{-0.24}\mathrm{exp}(-T/4000)$} & \cite{Stancil} \\
53  & \HeId+\HII+$\gamma$ $\rightarrow$ \HeHIId+$\gamma$, stimulated radiative association & \textbf{$3.2\times10^{-20}T^{1.8}/(1+0.1T^{2.04})*\mathrm{exp}(-T/4000)(1+2\times10^{-4}T_r^{1.1}) $} & JSK95, ZSD98 \\
54  & \HeId+\HzII $ \rightarrow$ \HeHIId+\HI & $3.0\times10^{-10}\mathrm{exp}(-6717/T)$ &\cite{Galli} \\
55  & \HeIId+\HI$\rightarrow$ \HeHIId+$\gamma$ & $4.16\times10^{-16}T^{-0.37}\mathrm{exp}(-T/87600)$ & SLD98 \\
56  & \HeHIId+\HI$\rightarrow$ \HeId+\HzII & $0.69\times10^{-9}(T/300)^{0.13}\mathrm{exp}(-T/33100)$ & LJB95 \\
57  & \HeHIId+\e$\rightarrow$ \HeId+\HI &$3.0\times10^{-8}(T/300)^{-0.47}$ & \cite{Stancil}\\
58  & \HeHIId+$\gamma$ $\rightarrow$ \HeId+\HII & $220T_r^{0.9}\mathrm{exp}(-22740/T_r)$ & JSK95 \\
59  & \HeHIId+$\gamma$ $\rightarrow$ \HeIId+\HI & $7.8\times10^3T_r^{1.2}\mathrm{exp}(-240000/T_r)$ & \cite{Galli} \\
\hline 
\end{longtable}
\end{landscape}
}% End \longtabL

\end{document}